\documentclass[preprint2]{proto}
\usepackage{times}
\usepackage{soul}

\newcommand{\Msun}{M_\odot}

\newcommand{\msun}{M_\odot}
\newcommand{\co}{\mbox{$^{12}$CO}}
\newcommand{\coa}{\mbox{$^{13}$CO}}

\usepackage{color}

\begin{document}
\title{\textbf{\LARGE Formation of Molecular Clouds and Global Conditions for Star Formation}}

\author {\textbf{\large Clare L. Dobbs}}
\affil{\small\em University of Exeter}

\author {\textbf{\large Mark R. Krumholz}}
\affil{\small\em University of California, Santa Cruz}

\author {\textbf{\large Javier Ballesteros-Paredes}}
\affil{\small\em Universidad Nacional Aut\'onoma de M\'exico}

\author {\textbf{\large Alberto D. Bolatto}}
\affil{\small\em University of Maryland, College Park}

\author {\textbf{\large Yasuo Fukui}}
\affil{\small\em Nagoya University}

\author {\textbf{\large Mark Heyer}}
\affil{\small\em University of Massachusetts}

\author {\textbf{\large Mordecai-Mark Mac Low}}
\affil{\small\em American Museum of Natural History}

\author {\textbf{\large Eve C. Ostriker}}
\affil{\small\em Princeton University}

\author {\textbf{\large Enrique V\'azquez-Semadeni}}
\affil{\small\em Universidad Nacional Aut\'onoma de M\'exico}

\begin{abstract}
\baselineskip = 11pt
\leftskip = 0.65in 
\rightskip = 0.65in
\parindent=1pc

{\small Giant molecular clouds (GMCs) are the primary reservoirs of cold, star-forming molecular gas in the Milky Way and similar galaxies, and thus any understanding of star formation must encompass a model for GMC formation, evolution, and destruction. These models are necessarily constrained by measurements of interstellar molecular and atomic gas, and the emergent, newborn stars. Both observations and theory have undergone great advances in recent years, the latter driven largely by improved numerical simulations, and the former by the advent of large-scale surveys with new telescopes and instruments. This chapter offers a thorough review of the current state of the field.
 \\~\\~\\~}

\end{abstract}

\section{\textbf{INTRODUCTION}}

\bigskip
\noindent

Stars form in a 
cold, dense, molecular phase of the interstellar medium (ISM) that  appears to be 
organized into coherent, localized volumes or clouds. The star formation history of the universe, the evolution of galaxies, and the formation
of planets in stellar environments are all coupled to the formation of these clouds, the collapse of unstable regions within
them to stars, and the clouds' final dissipation. The physics of these regions is complex, and descriptions of cloud structure and evolution remain incomplete and require continued exploration. Here we review the current status of observations and theory of molecular clouds, focusing on key advances in the field since Protostars and Planets V.

The first detections of molecules in the ISM date from the 1930s, with the discovery of CH and CN within the diffuse interstellar bands \citep{Swings:1937, Mckellar:1940} and later the microwave lines of OH \citep{Weinreb:1963}, NH$_3$ \citep{cheung:1968}, water vapor \citep{cheung:1969} and H$_2$CO \citep{Snyder:1969}. Progress accelerated in the 1970s with the first measurements of molecular hydrogen \citep{Carruthers:1970} and the \co\ J=1-0 line at 2.6mm \citep{Wilson:1970} and the continued development of millimeter wave instrumentation and facilities.  

The first maps of CO emission in nearby star forming regions and along the Galactic Plane revealed the unexpectedly large spatial extent of giant molecular clouds \citep[GMCs][]{Kutner:1977, Lada:1976, Blair:1978, Blitz:1980b}, and their substantial contribution to the mass budget of the ISM \citep{Scoville:1975, Gordon:1976, Burton:1978, Sanders:1984}. Panoramic imaging of \co\ emission in the Milky Way from both the Northern and Southern Hemispheres enabled the first complete view of the molecular gas distribution in the Galaxy \citep{Dame:1987, Dame:2001} and the compilation of GMC properties \citep{Solomon:1987,Scoville:1987}. Higher angular resolution observations of optically thin tracers of molecular gas in nearby clouds revealed a complex network of filaments \citep{Bally:1987, Heyer:1987}, and high density tracers such as NH$_3$, CS, and HCN revealed the dense regions of active star formation \citep{Myers:1983, Snell:1984}. Since this early work, large, millimeter filled aperture (IRAM~30m, NRO~45m), interferometric (BIMA, OVRO, Plateau de Bure) and submillimeter (CSO, JCMT) facilities have provided improved sensitivity and the ability 
to measure higher excitation
conditions.
Observations to date have identified $\sim$200 distinct interstellar molecules \citep{vanDishoeck:1998,Muller:2005}, and the last 40 years of observations using these molecules have determined a set of cloud properties on which our limited understanding of cloud physics is based. 

Theoretically, the presence of molecular hydrogen in the ISM was predicted long before the development of large scale CO surveys (e.g., \citealt{Spitzer:1949}). In the absence of metals, formation of H$_2$ 
   by gas phase reactions catalyzed by electrons and protons
is extremely slow, but dust grains catalyze the reaction and speed it up by orders of magnitude. As a result, H$_2$ formation is governed by the density of dust grains, gas density, and the ability of hydrogen atoms to stick to dust grains and recombine \citep{Hulst:1948,McCrea:1960,Gould:1963,Hollenbach:1971}. The ISM exhibits a sharp transition in molecular fraction from low to high densities, typically at 1--100 cm$^{-3}$ (or $\Sigma\sim$1--100 M$_{\odot}$ pc$^{-2}$), dependent mostly on the UV radiation field and metallicity \citep{vanDishoeck:1986, Pelupessy:2006, Glover:2007a, Dobbs:2008c, Krumholz:2008, Krumholz:2009c, Gnedin:2009}. This dramatic increase in H$_2$ fraction represents a 
change to the regime where H$_2$ becomes self shielding. Many processes have been invoked to explain how atomic gas reaches the densities ($\gtrsim 100$ cm$^{-3}$) required to become predominantly molecular (see Section~3). Several mechanisms likely to govern ISM structure became apparent in the 
1960s: cloud-cloud collisions \citep{Oort:1954,Field:1965b}, gravitational instabilities (e.g., \citealt{Lynden-Bell:1965}), thermal instabilities \citep{Field:1965a}, and magnetic instabilities \citep{Parker:1966,Mouschovias:1974}. At about the same time, \citet{Roberts:1969} showed that the gas response to a stellar spiral arm produces a strong spiral shock, likely observed as dust lanes and associated with molecular gas. Somewhat more recently, the idea of cloud formation from turbulent flows in the ISM has emerged \citep{Ballesteros:1999}, as well as colliding flows from stellar feedback processes \citep{Koyama:2000}. 

The nature of GMCs, their lifetime, and whether they are virialized, remains
unclear. Early models of cloud-cloud collisions required very long-lasting clouds (100 Myr) in order to build up more massive GMCs \citep{Kwan:1979}. Since then, lifetimes have generally been revised downwards. Several observationally derived estimates, including up to the present date, have placed cloud lifetimes at around 20--30~Myr
\citep{Bash:1977,Blitz:1980,Fukui:1999,Kawamura:2009,Miura:2012}, although there have been longer estimates for molecule rich galaxies \citep{Tomisaka:1986, Koda:2009} and shorter
estimates for smaller, nearby clouds \citep{Elmegreen:2000,Hartmann:2001}. 

In the 1980s and 1990s,
GMCs were generally thought to be supported against gravitational collapse, and in virial equilibrium. Magnetic fields were generally favored as a means of support \citep{Shu:1987}. Turbulence would dissipate unless replenished, whilst rotational support was found to be insufficient (e.g., \citealt{Silk:1980}). 
More recently these conclusions have been challenged by new observations, which have revised estimates of magnetic field strengths downwards, and new simulations and theoretical models that suggest that clouds may in fact be turbulence-supported, or that they may be entirely transient objects that are not supported against collapse at all. These questions are all under active discussion, as we review below.

In \S~2, we describe the main new observational results, and corresponding theoretical interpretations. These include the extension of the Schmidt-Kennicutt relation to other tracers, notably H$_2$, as well as to much smaller scales, e.g., those of individual clouds. \S~2 also examines the latest results on GMC properties, both within the Milky Way and in external galaxies. Compared to the data that were available at the time of PPV, CO surveys offer much higher resolution and sensitivity within the Milky Way, and are able to better cover a wider range of environments beyond it. In \S~3, we discuss GMC formation, providing a summary of the main background and theory, whilst reporting the main advances in numerical simulations since PPV. We also discuss progress on calculating the conversion of atomic to molecular gas, and CO chemistry.
\S~4 describes the various scenarios for the evolution of GMCs, including the revival of globally collapsing clouds  as a viable theoretical model, 
and examines the role of different forms of stellar feedback as internal and external sources of cloud motions. Then in \S~5 we relate the star forming properties in GMCs to these different scenarios. Finally in \S~6 we look forward to what we can expect between now and PPVII.

\bigskip

\centerline{\textbf{ 2. OBSERVED PROPERTIES OF GMCs}}
\bigskip

\noindent
\textbf{2.1. GMCs and Star Formation}
\bigskip

Molecular gas is strongly correlated with star formation on scales from entire galaxies \citep{Kennicutt:1989, Kennicutt:1998, Gao:2004, Saintonge:2011} to kpc and sub-kpc regions \citep{Wong:2002, Bigiel:2008,Rahman:2012,Leroy:2013} to individual GMCs \citep{Evans:2009,Heiderman:2010,Lada:2010,Lada:2012}. These relations take on different shapes at
different scales. Early studies of whole galaxies found a power-law correlation between total gas content (H~\textsc{i} plus H$_2$) and star formation rate (SFR) with an index $N\sim1.5$ \citep{Kennicutt:1989, Kennicutt:1998}. These studies include galaxies that span a very large range of properties, from dwarfs to ultraluminous IR galaxies, so it is possible that
the physical underpinnings of this relation are different in different regimes. Transitions with higher critical densities such as HCN($1-0$) and higher-$J$ CO lines \citep{Gao:2004, Bayet:2009, Juneau:2009, Garcia-Burillo:2012} also show power-law correlations but with smaller indices; the index appears to depend mostly on the line critical density, a result that can be explained through models \citep{Krumholz:2007a, Narayanan:2008, Narayanan:2008a}.

Within galaxies the star formation rate surface density, $\Sigma_{SFR}$, is strongly correlated with the surface density of molecular gas as traced by CO emission and only very weakly, if at all, related to atomic gas. The strong correlation with H$_2$ persists even in regions where atomic gas dominates the mass budget \citep{Schruba:2011,Bolatto:2011}. The precise form of the SFR--H$_2$ correlation is a subject of study, with results spanning the range from super-linear \citep{Kennicutt:2007,Liu:2011,Calzetti:2012} to approximately linear \citep{Bigiel:2008,Blanc:2009,Rahman:2012,Leroy:2013b} to sub-linear \citep{Shetty:2013}. Because CO is used to trace H$_2$, the correlation
can be altered by systematic variations in the CO to H$_2$ conversion factor, an effect that may flatten the observed relation compared to the
true one \citep{Shetty:2011,Narayanan:2011,Narayanan:2012}.

The SFR--H$_2$ correlation defines a molecular depletion time, $\tau_{\rm dep}({\rm H_2})=M({\rm H}_2)/{\rm SFR}$, which is the time required to consume all the H$_2$ at the current SFR. 
A linear SFR--H$_2$ correlation implies a constant $\tau_{\rm dep}({\rm H_2})$, while super-linear (sub-linear) relations yield a time scale 
$\tau_{\rm dep}({\rm H_2})$ that decreases (increases) with surface density. In regions where CO emission is present, the mean depletion time over kpc scales is $\tau_{\rm dep}({\rm H_2})=2.2$ Gyr with $\pm0.3$ dex scatter, with some dependence on the local conditions \citep{Leroy:2013b}. \citet{Saintonge:2011b} find that, for entire galaxies, $\tau_{\rm dep}({\rm H_2})$ decreases by a factor of $\sim3$ over two orders of magnitude increase in the SFR surface. \citet{Leroy:2013b} show that the kpc-scale measurements within galaxies are consistent with this trend, but that $\tau_{\rm dep}({\rm H_2})$ also correlates
with the dust-to-gas ratio. For normal galaxies, using a CO-to-H$_2$ conversion factor that depends on the local dust-to-gas ratio removes most of the variation in $\tau_{\rm dep}({\rm H_2})$.

On scales of a few hundred parsecs, the scatter in $\tau_{\rm dep}({\rm H_2})$ rises significantly \citep[e.g.,][]{Schruba:2010,Onodera:2010} and the SFR--H$_2$ correlation breaks down. This is partially a manifestation of the large dispersion in SFR per unit mass in individual GMCs \citep{Lada:2010}, but it is also a consequence of the time scales involved \citep{Kawamura:2009, Kim:2013}. Technical issues concerning the interpretation of the tracers also become important on the small scales \citep{Calzetti:2012}. 

On sub-GMC scales there are strong correlations between star formation and extinction, column density, and volume density. The correlation with volume density is very close to that observed in 
ultraluminous IR galaxies \citep{Wu:2005}. Some authors have interpreted these data as implying that star formation only begins above a threshold column density of $\Sigma_{\rm H_2}\sim$110--130 M$_\odot$~yr$^{-1}$ or volume density $n \sim 10^{4-5}$ cm$^{-3}$ \citep{Evans:2009,Heiderman:2010,Lada:2010,Lada:2012}. However, others argue that the data are equally consistent with a smooth rise in SFR with volume or surface density, without any particular threshold value \citep{Krumholz:2007, Narayanan:2008, Narayanan:2008a, Gutermuth:2011, Krumholz:2012, Burkert:2013}.

\bigskip
\noindent
\textbf{2.2. GMCs as a Component of the Interstellar Medium}
\bigskip
 
Molecular clouds are the densest, coldest, highest column density, highest extinction component of the interstellar medium. Their masses are dominated by molecular gas (H$_2$), with a secondary contribution from He ($\sim26\%$), and a varying contribution from H~\textsc{i} in a cold envelope \citep[e.g.,][]{Fukui:2010} and interclump gas detectable by H~{\sc i} self-absorption \citep{Goldsmith:2005}. 
Most of the molecular mass in galaxies is in the form of molecular clouds,
with the possible exception of galaxies with gas surface densities substantially higher than that of the Milky Way, where a substantial diffuse H$_2$ component exists \citep{Papadopoulos:2012,Papadopoulos:2012a, Pety:2013, Colombo:2013}.

Molecular cloud
masses range from $\sim 10^2$ $M_\odot$ for small clouds at high Galactic latitudes \citep[e.g.,][]{Magnani:1985} and in the outer disk of the Milky Way \citep[e.g.,][]{Brand:1995,Heyer:2001} up to giant $\sim 10^7$ $M_\odot$ clouds in the central molecular zone of the Galaxy \citep{Oka:2001}. 
The measured mass spectrum 
of GMCs (see \S2.3) implies
that most of the molecular mass resides in the largest GMCs. Bulk densities of clouds are $\log[n_{\rm H_2}/{\rm cm^{-3}}]=2.6\pm0.3$ \citep{Solomon:1987, RomanDuval:2010}, but clouds have inhomogenous density distributions with large contrasts \citep{Stutzki:1988}.  The ratio of molecular to stellar mass in galaxies shows a
strong trend with galaxy color from high in blue galaxies (10\% for $NUV-r\sim2$) to low in red galaxies ($\lesssim0.16\%$ for $NUV-r\gtrsim5$) \citep{Saintonge:2011}. The typical molecular to atomic ratio in galaxies where both H~\textsc{i} and H$_2$ are detected is ${R_{\rm mol}\equiv M_{\rm H_2}/M_{\rm HI}\approx0.3}$ with scatter of $\pm0.4$ dex. The large scatter reflects the fact that the atomic and molecular masses are only weakly correlated, and in contrast with the molecular gas to stellar mass fraction, 
the ratio $R_{\rm mol}$ shows only weak correlations with galaxy properties such as color
\citep{Leroy:2005,Saintonge:2011}. 

\begin{figure*}
\epsscale{1.7}
\plotone{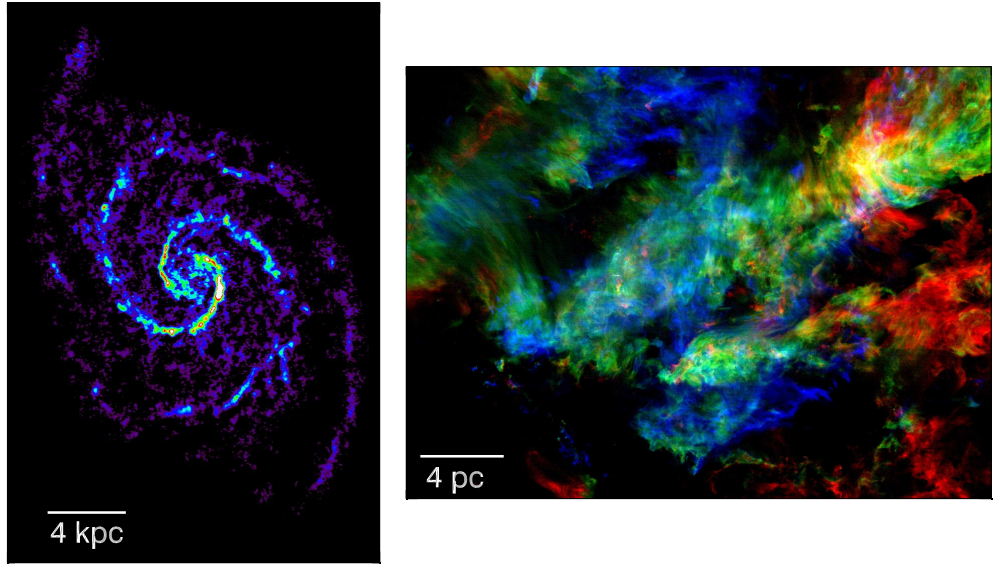}
\caption{\small
\label{fig:m51_taurus}
(left) CO J=1-0 image of M51 from \citet{Koda:2009} showing the largest cloud complexes are distributed in spiral arms, while smaller GMCs lie
both in and between spiral features. 
(right) 3 color image of CO J=1-0 emission from the Taurus molecular cloud from 
\citet{Narayanan:2008b} illustrating complex gas motions within clouds. Colors represents the CO integrated intensities
over $V_{LSR}$ intervals 0-5 (blue), 5-7.5 (green) and 7.5-12 (red) km s$^{-1}$.
}
\end{figure*}

In terms of their respective spatial distributions, in spiral galaxies H$_2$ is reasonably well described by an exponential profile with a scale length
    $\ell_{\rm CO} \approx 0.2\,R_{25}$, rather smaller than
the optical emission
\citep{Young:1995,Regan:2001,Leroy:2009,Schruba:2011}, where 
$R_{25}$ is the 25th magnitude isophotal radius for the stellar light distribution. In contrast,  H~\textsc{i} shows a nearly flat distribution with typical maximum surface density $\Sigma_{\rm HI,max}\sim12$~M$_\odot\,{\rm pc}^{-2}$ (similar to the H~\textsc{i} column seen toward Solar neighborhood clouds, \citealt{Lee:2012}). Galaxy centers are the regions that show the most variability and the largest departures from these trends \citep{Regan:2001, Bigiel:2012}. At low metallicities the H~\textsc{i} surface density can be much larger, possibly scaling as $\Sigma_{\rm HI,max} \sim Z^{-1}$ 
\citep{Fumagalli:2010,Bolatto:2011,Wong:2013}. In spiral galaxies the transition between the atomic- and molecular-dominated regions occurs at $R\sim0.4\,R_{25}$ \citep[e.g.,][]{Leroy:2008}. The CO emission also shows much more structure than the H~\textsc{i} on the small scales \citep{Leroy:2013b}. In spirals with well
defined arms (NGC 6946, M~51, NGC628) the interarm regions contain at least 30\% of the measured CO luminosity \citep{Foyle:2010}, but at fixed total gas surface density $R_{\rm mol}$ is very similar for arm and interarm regions, suggesting that arms act mostly to collect gas rather to directly trigger H$_2$
formation \citep{Foyle:2010} (see Figure~\ref{fig:m51_taurus}). We discuss the relationship between H~\textsc{i} and H$_2$ in more detail in \S~3.3.

\bigskip
\noindent
\textbf{ 2.3. Statistical Properties of GMCs}
\bigskip
\label{sec:GMCstatistics}

Statistical descriptions of GMC properties have provided insight into the processes that govern their formation and evolution since large surveys first became possible in the 1980s (see \S1).
While contemporary observations are more sensitive and feature better angular resolution and sampling than earlier surveys, identification of clouds within position-position-velocity (PPV) data cubes remains a significant problem. In practice, one defines a cloud as a set of contiguous voxels in a PPV data cube of CO emission above a surface brightness threshold. Once a cloud is defined, one can compute global properties such as size, velocity dispersion, and luminosity \citep{Williams:1994, Rosolowsky:2006}.  While these algorithms
have been widely applied, their reliability and completeness are difficult to evaluate 
\citep{Ballesteros:2002,Pineda:2009,Kainulainen:2009b}, 
particularly for surveys of \co\ and \coa\ in the Galactic Plane that are subject to blending of emission from unrelated clouds. The improved resolution of modern surveys helps reduce these problems, but higher surface brightness thresholds are required to separate a feature in velocity-crowded regions. High resolution can also complicate the accounting, as the algorithms may identify cloud substructure as distinct clouds. Moreover, even once a cloud is identified, deriving masses and mass-related quantities from observed CO emission generally requires application of the CO-to-H$_2$ conversion factor or the H$_2$ to $^{13}$CO abundance ratio, both of which can vary within and between clouds in response to local conditions of UV irradiance, density, temperature, and metallicity \citep{Bolatto:2013, Ripple:2013}. Millimeter wave interferometers can resolve large GMC complexes in nearby galaxies but must also account for missing flux from an extended component of emission.

Despite these observational difficulties, there are some robust results. Over the mass range $M>10^4$ $M_\odot$ where it can be measured reliably, the cloud mass spectrum is well-fit by a powerlaw $dN/dM \sim M^{-\gamma}$ (cumulative distribution function $N(>M) \sim M^{-\gamma+1}$), with values $\gamma < 2$ indicating that most of the mass is in large clouds. For GMCs in the Milky Way, $\gamma$ is consistently found to be in the range $1.5$ to $1.8$ \citep{Solomon:1987, Kramer:1998, Heyer:2001, RomanDuval:2010} 
with the higher value likely biased by the inclusion of cloud fragments identified as distinct clouds. GMCs in the Magellanic Clouds exhibit a steeper mass function overall and specifically for massive clouds \citep{Fukui:2008, Wong:2011}. In M33, $\gamma$ ranges from $1.6$ in the inner regions to $2.3$ at larger radii  \citep{Rosolowsky:2005, Gratier:2012}. 

In addition to clouds' masses, we can measure their sizes and thus their surface densities. The \citet{Solomon:1987} catalog of inner Milky Way GMCs, updated to the current Galactic distance scale, shows a distribution of GMCs surface densities $\Sigma_{\rm GMC}\approx150^{+95}_{-70}$~\mbox{${\rm M}_\odot
{\rm pc}^{-2}$} ($\pm1\sigma$ interval) assuming a fixed CO-to-H$_2$ conversion factor $X_{\rm CO}=2\times10^{20}$ cm$^{-2}$ (K km s$^{-1}$)$^{-1}$, and including the He mass \citep{Bolatto:2013}.  
\citet{Heyer:2009} re-observed these clouds in $^{13}$CO and found $\Sigma_{\rm GMC}\sim40$~\mbox{${\rm M}_\odot {\rm pc}^{-2}$} over the same cloud areas, but concluded that this is likely at least a factor of 2 too low due to non-LTE and optical depth effects. \citet{Heiderman:2010} find that
$^{13}$CO can lead to a factor of 5 underestimate. A reanalysis by \citet{RomanDuval:2010} shows $\Sigma_{\rm GMC}\sim144$~\mbox{${\rm M}_\odot {\rm pc}^{-2}$} using the $^{13}$CO rather than the $^{12}$CO contour to define the area. 
Measurements of surface densities in extragalactic GMCs remain challenging, but with the advent of ALMA the field is likely to evolve quickly. For a sample of nearby galaxies, many of them dwarfs, \citet{Bolatto:2008} find $\Sigma_{\rm GMC}\approx85$ M$_\odot$~pc$^{-2}$. Other recent extragalactic surveys find roughly comparable results, $\Sigma_{\rm GMC}\sim40-170$ M$_\odot$~pc$^{-2}$ \citep{Rebolledo:2012,Donovan:2013}.

GMC surface densities may prove to be a function of environment. The PAWS survey of M~51 finds a progression in surface density \citep{Colombo:2013}, from clouds in the center ($\Sigma_{\rm GMC}\sim210$~\mbox{${\rm M}_\odot {\rm pc}^{-2}$}), to clouds in arms ($\Sigma_{\rm GMC}\sim185$~\mbox{${\rm M}_\odot {\rm pc}^{-2}$}), to those in interarm regions ($\Sigma_{\rm GMC}\sim140$~\mbox{${\rm M}_\odot {\rm pc}^{-2}$}). \citet{Fukui:2008}, \citet{Bolatto:2008}, and \citet{Hughes:2010} find that GMCs in the Magellanic Clouds have lower surface densities than those in the inner Milky Way ($\Sigma_{\rm GMC}\sim50$~\mbox{${\rm M}_\odot {\rm pc}^{-2}$}). Because of the presence of extended H$_2$ envelopes at low metallicities (\S2.6), however, this may underestimate their true molecular surface density \citep[e.g.,][]{Leroy:2009}. Even more extreme variations in $\Sigma_{\rm GMC}$ are observed near the Galactic Center and in more extreme starburst environments (see \S~2.7).

In addition to studying the mean surface density of GMCs, observations within the Galaxy can also probe the distribution of surface densities within GMCs. For a sample of Solar neighborhood clouds, \citet{Kainulainen:2009} use infrared extinction measurements to determine that PDFs of column densities are lognormal from $0.5 < A_V < 5$ (roughly 10--100~$\msun$ pc$^{-2}$), with a power-law tail at high column densities in actively star-forming clouds. Column density images derived from dust emission also find such excursions \citep{Schneider:2012,Schneider:2013}. \citet{Lombardi:2010}, also using infrared extinction techniques, find that, although GMCs contain a wide range of column densities, the mass $M$ and area $A$ contained within a specified extinction threshold nevertheless obey the \citet{Larson:1981} $M\propto A$ relation, which implies constant column density. 

Finally, we warn that \textit{all} column density measurements are subject to a potential bias.  GMCs are identified as contiguous areas with surface brightness values or extinctions above a threshold typically set by the sensitivity of the data.  Therefore, pixels at or just above this threshold comprise most of the area of the defined cloud and the measured cloud surface density is likely biased towards the column density associated with this threshold limit. Note that there is also a statistical difference between ``mass-weighed" and ``area-weighed" $\Sigma_{\rm GMC}$. The former is the average surface density that contributes most of the mass, while the latter represents a typical surface density over most of the cloud extent. Area-weighed $\Sigma_{\rm GMC}$ tend to be lower, and although perhaps less interesting from the viewpoint of star formation, they are also easier to obtain from observations.

\begin{figure}[htb]
\epsscale{1.0}
\plotone{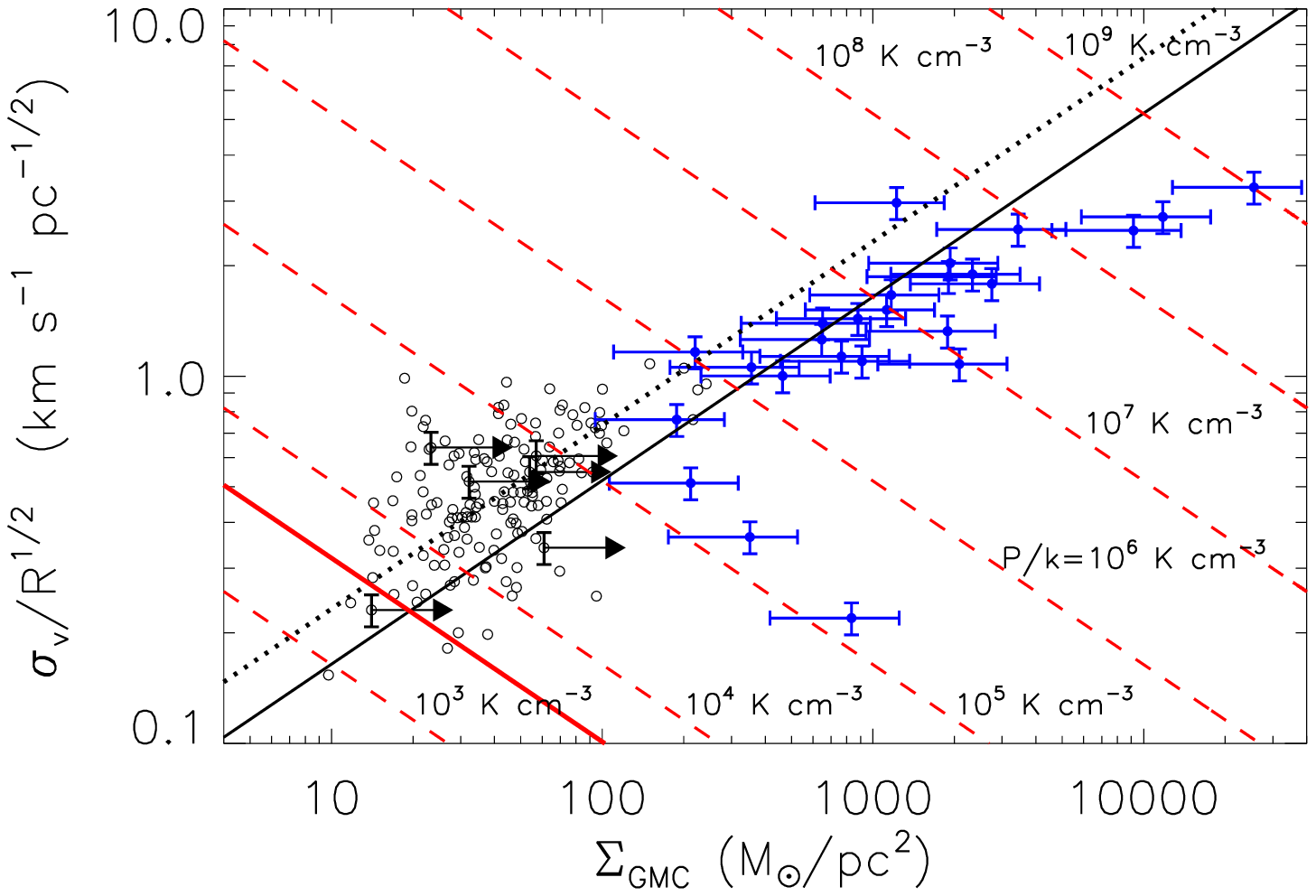}
\caption{\small
\label{fig:heyer_larson}
The variation of $\sigma_v/R^{1/2}$ with surface density, $\Sigma_{\rm GMC}$, for Milky Way GMCs from \citet{Heyer:2009} (open circles) and massive cores from \citet{Gibson:2009} (blue points).  
For clarity, a limited number of error bars are displayed for the GMCs.  The horizontal error bars for the GMCs convey lower limits to the mass surface density derived from \coa.  The vertical error bars for both data sets reflect a 20\% uncertainty in the kinematic distances.  The horizontal error bars for the massive cores assume a 50\% error in the C$^{18}$O and N$_2$H$^+$ abundances used to derive mass. 
The solid and dotted black lines show loci corresponding to gravitationally bound and marginally bound clouds respectively.  Lines of constant turbulent pressure are illustrated by the red dashed lines.  The mean thermal pressure of the local ISM is shown as the red solid line.
}
\end{figure}

In addition to mass and area, velocity dispersion is the third quantity that we can measure for a large sample of clouds.  It provides a coarse assessment of the complex motions in GMCs as illustrated in Figure~\ref{fig:m51_taurus}.
\citet{Larson:1981} identified scaling relationships between velocity dispersion and cloud size suggestive of a turbulent velocity spectrum, and a constant surface density for clouds.   Using more sensitive surveys of GMCs, 
\citet{Heyer:2009} found a scaling relation that extends the Larson relationships such that the one-dimensional velocity dispersion $\sigma_v$ depends 
on the physical radius, $R$, and the column density $\Sigma_{\rm GMC}$, as shown in Figure~\ref{fig:heyer_larson}.
The points follow the expression, 
\begin{equation}
\sigma_v = 0.7 (\Sigma_{\rm GMC}/100 M_\odot\,\rm{pc}^{-2})^{1/2}  (R/1\,{\rm pc})^{1/2}\mbox{ km s}^{-1}.
\label{eq:heyer}
\end{equation}
More recent compilations of GMCs in the Milky Way \citep{RomanDuval:2010} have confirmed this result, and studies of Local Group galaxies \citep{Bolatto:2008, Wong:2011} have shown that it applies to GMCs outside the Milky Way as well.  
Equation~\ref{eq:heyer} is a natural consequence of gravity playing an important role in setting the characteristic velocity scale in clouds, either through collapse \citep{Ballesteros:2011a} or virial equilibrium \citep{Heyer:2009}.
Unfortunately one expects only factor of $\sqrt{2}$ differences in velocity dispersion between clouds that are in free-fall collapse or
in virial equilibrium \citep{Ballesteros:2011a} 
making it extremely difficult to distinguish between these possibilities using observed scaling relations. 
Concerning the possibility of pressure-confined but mildly self-gravitating clouds 
\citep{Field:2011}, Figure \ref{fig:heyer_larson} shows that the turbulent pressures, $P=\rho \sigma_v^2$, in observed GMCs are generally larger than the mean thermal pressure of the diffuse ISM \citep{Jenkins:2011} so these structures must be confined by self-gravity.  

As with column density, observations within the Galaxy can also probe internal velocity structure. \citet{Brunt:2003}, \citet{Heyer:2004}, and \citet{Brunt:2009} used principal components analysis of GMC velocity fields to investigate the scales on which turbulence in molecular clouds could be driven. They found no break in the velocity dispersion-size relation, and reported that the second principle component has a ``dipole-like'' structure. Both features suggest that the dominant processes driving GMC velocity structure must operate on scales comparable to or larger than single clouds.

\bigskip
\noindent
\textbf{ 2.4. Dimensionless Numbers: Virial Parameter and Mass to Flux Ratio}
\bigskip

The virial theorem describes the large-scale dynamics of gas in GMCs, so ratios of the various terms that appear in it are 
a useful guide to what forces are important in GMC evolution. Two of these ratios are the virial parameter, which evaluates the importance of internal pressure and bulk motion relative to gravity, and the dimensionless mass to flux ratio, which describes the importance of magnetic fields compared to gravity. Note, however, that neither of these ratios accounts for potentially-important surface terms \citep[e.g.,][] {Ballesteros:1999}.

The virial parameter is defined as $\alpha_G=M_{\rm virial}/M_{\rm GMC}$, where $M_{\rm virial}=5\sigma_v^2 R/G$ and $M_{\rm GMC}$ is the luminous mass of the cloud. For a cloud of uniform density with negligible surface pressure and magnetic support, $\alpha_G = 1$ corresponds to virial equilibrium and $\alpha_G = 2$ to being marginally gravitationally bound, although in reality $\alpha_G > 1$ does not strictly imply expansion, nor does $\alpha_G <1$ strictly imply contraction \citep{Ballesteros:2006a}. Surveys of the Galactic Plane and nearby galaxies using \co\ emission to identify clouds find an excellent, near-linear correlation between $M_{\rm virial}$ and the CO luminosity, $L_{\rm CO}$, with a coefficient implying that (for reasonable CO-to-H$_2$ conversion factors) the typical cloud virial parameter is unity \citep{Solomon:1987, Fukui:2008, Bolatto:2008, Wong:2011}. Virial parameters for clouds exhibit a range of values from $\alpha_G \sim 0.1$ to $\alpha_G \sim 10$, but typically $\alpha_G$ is indeed $\sim 1$.
\citet{Heyer:2009} reanalyzed the \citet{Solomon:1987} GMC sample using $^{13}$CO J=1-0 emission to derive cloud mass and found a median $\alpha_G=1.9$. This value is still consistent with a median $\alpha_G = 1$, since excitation and abundance variations in the survey lead to systematic underestimates of $M_{\rm GMC}$. A cloud catalog generated directly from the \coa\ emission of the BU-FCRAO Galactic Ring Survey resulted in a median $\alpha_G=0.5$ \citep{RomanDuval:2010}. Previous surveys \citep{Dobashi:1996,Yonekura:1997,Heyer:2001} tended to find higher $\alpha_G$ for low mass clouds, possibly a consequence of earlier cloud-finding algorithms preferentially decomposing single GMCs into smaller fragments \citep{Bertoldi:1992}. 

The importance of magnetic forces is characterized by the ratio $M_{\rm GMC}/M_{\rm cr}$, where $M_{\rm cr}=\Phi/(4\pi^2G)^{1/2}$ and $\Phi$ is the magnetic flux threading the cloud \citep{Mouschovias:1976, Nakano:1978}. If $M_{\rm GMC}/M_{\rm cr}>1$ (the supercritical case) then the magnetic field is incapable of providing the requisite force to balance self-gravity, while if $M_{\rm GMC}/M_{\rm cr}<1$ (the subcritical case) the cloud can be supported against self-gravity by the magnetic field. Initially subcritical volumes can become supercritical through ambipolar diffusion \citep{Mouschovias:1987, Lizano:1989}. Evaluating whether a cloud is sub- or supercritical is challenging. Zeeman measurements of the OH and CN lines offer a direct measurement of the line of sight component of the magnetic field at densities $\sim 10^3$ and $\sim 10^5$ cm$^{-3}$, respectively, but statistical corrections are required to account for projection effects for both the field and the column density distribution. \citet{Crutcher:2012} provides a review of techniques and observational results, and report a mean value $M_{\rm GMC}/M_{\rm cr} \approx 2-3$, implying that clouds are generally supercritical, though not by a large margin.

\bigskip
\noindent
\textbf{2.5. GMC Lifetimes}
\bigskip

The natural time unit for GMCs is the free-fall time, which for a medium of density $\rho$ is given by $\tau_{\rm ff} = [3\pi/(32 G \rho)]^{1/2} = 3.4 (100/n_{\rm H_2})^{1/2}$ Myr, where $n_{\rm H_2}$ is the number density of H$_2$ molecules, and the mass per H$_2$ molecule is $3.9\times 10^{-24}$ g for a fully molecular gas of cosmological composition. This is the timescale on which an object that experiences no significant forces other than its own gravity will collapse to a singularity. For an object with $\alpha_G \approx 1$, the crossing timescale is $\tau_{\rm cr} = R/\sigma \approx 2\tau_{\rm ff}$. It is of great interest how these natural timescales compare to cloud lifetimes and depletion times.

\citet{Scoville:1979} argue that GMCs in the Milky Way are very long-lived ($>10^8$ yr) based on the detection of molecular clouds in interarm regions, and \citet{Koda:2009} apply similar arguments to the H$_2$-rich galaxy M51. They find that, while the largest GMC complexes reside within the arms, smaller ($<10^4$ $\msun$) clouds are found in the interarm regions, and the molecular fraction is large ($>75\%$) throughout the central 8~kpc \citep[see also][]{Foyle:2010}. This suggests that massive GMCs are rapidly built-up in the arms from smaller, pre-existing clouds that survive the transit between spiral arms. The massive GMCs fragment into the smaller clouds upon exiting the arms, but have column densities high enough to remain molecular (see \S~3.4). Since the time between spiral arm passages is $\sim 100$ Myr, this implies a similar cloud lifetime $\tau_{\rm life} \gtrsim 100\mbox{ Myr} \gg \tau_{\rm ff}$. Note, however, this is an argument for the mean lifetime of a H$_2$ molecule, not necessarily for a single cloud. Furthermore, these arguments do not apply to H$_2$-poor galaxies like the LMC and M33.

\citet[][see also \citealt{Fukui:1999,Gratier:2012}]{Kawamura:2009} 
use the NANTEN Survey of \co\ J=1-0 emission from the LMC, which is complete for clouds with mass $>5\times 10^4$ $\msun$, to identify three distinct cloud types that are linked to specific phases of cloud evolution.  Type I clouds are devoid of massive star formation and represent the earliest phase. Type II clouds contain compact H~\textsc{ii} regions, signaling the onset of massive star formation. Type III clouds, the final stage, harbor developed stellar clusters and H~\textsc{ii} regions. The number counts of cloud types indicate the relative lifetimes of each stage, and age-dating the star clusters found in type III clouds then makes it possible to assign absolute durations of 6, 13, and 7 Myrs for Types I, II, and III respectively. Thus the cumulative GMC lifetime is $\tau_{\rm life} \sim 25$ Myrs. This is still substantially greater than $\tau_{\rm ff}$, but by less so than in M51.

While lifetime estimates in external galaxies are possible only for large clouds, in the Solar Neighborhood it is possible to study much smaller clouds, and to do so using timescales derived from the positions of individual stars on the HR diagram. \citet{Elmegreen:2000, Hartmann:2001} and \citet{Ballesteros:2007b}, examining a sample of Solar Neighborhood GMCs, note that their HR diagrams are generally devoid of post T-Tauri stars with ages of $\sim 10$ Myr or more, suggesting this as an upper limit on $\tau_{\rm life}$. More detailed analysis of HR diagrams, or other techniques for age-dating stars, generally points to age spreads of at most $\sim 3$ Myr \citep{Reggiani:2011, Jeffries:2011}.

While the short lifetimes inferred for Galactic clouds might at first seem inconsistent
with the extragalactic data, it is important to remember that the two data sets are probing essentially non-overlapping ranges of cloud mass and length scale. The largest Solar Neighborhood clouds that have been age-dated via HR diagrams have masses $< 10^4$ $\msun$ (the entire Orion cloud is more massive than this, but the age spreads reported in the literature are only for the few thousand $M_\odot$ central cluster), below the detection threshold of most extragalactic surveys. Since larger clouds have, on average, lower densities and longer free-fall timescales, the difference in $\tau_{\rm life}$ is much larger than the difference in $\tau_{\rm life} / \tau_{\rm ff}$. Indeed, some authors argue that $\tau_{\rm life}/\tau_{\rm ff}$ may be $\sim 10$ for Galactic clouds as well as extragalactic ones \citep{Tan:2006}.

\bigskip
\noindent
\textbf{2.6. Star Formation Rates and Efficiencies}
\bigskip

We can also measure star formation activity within clouds. We define the star formation efficiency or yield, $\epsilon_*$, as the instantaneous fraction of a cloud's mass that has been transformed into stars, $\epsilon_* = M_*/(M_*+M_{\rm gas})$, where $M_*$ is the mass of newborn stars. In an isolated, non-accreting cloud, $\epsilon_*$ increases monotonically, but in an accreting cloud it can decrease as well. \citet{Krumholz:2005}, building on classical work by \citet{Zuckerman:1974b}, argue that a more useful quantity than $\epsilon_*$ is the star formation efficiency per free-fall time, defined as $\epsilon_{\rm ff} = \dot{M}_* / (M_{\rm gas}/\tau_{\rm ff})$, where $\dot{M}_*$ is the instantaneous star formation rate. This definition can also be phrased in terms of the depletion timescale introduced above: $\epsilon_{\rm ff} = \tau_{\rm ff}/\tau_{\rm dep}$. One virtue of this definition is that it can be applied at a range of densities $\rho$, by computing $\tau_{\rm ff}(\rho)$ then taking $M_{\rm gas}$ to be the mass at a density $\geq \rho$ \citep{Krumholz:2007}. 
As newborn stars form in the densest regions of 
clouds, $\epsilon_*$ can only increase as one increases the density threshold used to define $M_{\rm gas}$.  
It is in principle possible for
$\epsilon_{\rm ff}$ to both increase and decrease, and its behavior as a function of density encodes important information about how star formation behaves.

Within individual clouds, the best available data on $\epsilon_*$ and $\epsilon_{\rm ff}$ come from campaigns that use the Spitzer Space Telescope to obtain a census of young stellar objects with excess infrared emission, a feature that persists for 2--3~Myr of pre-main sequence evolution. These are combined with cloud masses and surface densities measured by millimeter dust emission or infrared extinction of background stars. For a set of five star forming regions investigated in the {\it Cores to Disks} Spitzer Legacy program, \citet{Evans:2009} found $\epsilon_*=$0.03--0.06 over entire GMCs, and $\epsilon_* \sim 0.5$ considering only dense gas with $n\sim 10^5$ cm$^{-3}$. On the other hand, $\epsilon_{\rm ff} \approx 0.03-0.06$ regardless of whether one considers the dense gas or the diffuse gas, due to a rough cancellation between the density dependence of $M_{\rm gas}$ and $\tau_{\rm ff}$. \citet{Heiderman:2010} obtain comparable values in 15 additional clouds from the Gould's Belt Survey. 
\citet{Murray:2011} find significantly higher values of $\epsilon_{\rm ff} = 0.14-0.24$ for the star clusters in the Galaxy that are brightest in WMAP free-free emission, but this value may be biased high because it is based on the assumption that the molecular clouds from which those clusters formed have undergone negligible mass loss despite the clusters'
 extreme luminosities \citep{Feldmann:2011}.

At the scale of the Milky Way as a whole, recent estimates based on a variety of indicators put the galactic star formation rate at $\approx 2$ $M_\odot$ yr$^{-1}$ \citep{Robitaille:2010, Murray:2010a, Chomiuk:2011}, 
within a factor of $\sim 2$ of earlier estimates based on ground-based radio catalogs \citep[e.g.,][]{Mckee:1997}. In comparison, the total molecular mass of the Milky Way is roughly $10^9$ $M_\odot$ \citep{Solomon:1987}, and this, combined with the typical free-fall time estimated in the previous section, gives a galaxy-average $\epsilon_{\rm ff} \sim 0.01$ \citep[see also][]{Krumholz:2007, Murray:2010a}.

For extragalactic sources one can measure $\epsilon_{\rm ff}$ by combining SFR indicators such as H$\alpha$, ultraviolet, and infrared emission with tracers of gas at a variety of densities. As discussed above, observed H$_2$ depletion times are $\tau_{\rm dep}({\rm H}_2) \approx 2$ Gyr, whereas GMC densities of $n_{\rm H} \sim $30--1000~cm$^{-3}$ correspond to free-fall times of $\sim$ 1--8~Myr, with most of the mass probably closer to the smaller value, since the mass spectrum of GMCs ensures that most mass is in large clouds, which tend to have lower densities. Thus $\epsilon_{\rm ff} \sim$0.001--0.003. 
Observations using tracers of dense gas ($n\sim 10^5$ cm$^{-3}$) such as HCN yield $\epsilon_{\rm ff} \sim 0.01$ \citep{Krumholz:2007, Garcia-Burillo:2012}; given the errors, the difference between the HCN and CO values is not significant. 
As with the \citet{Evans:2009} clouds, higher density regions subtend  smaller volumes and comprise smaller masses. $\epsilon_{\rm ff}$ is nearly constant because $M_{\rm gas}$ and $1/\tau_{\rm ff}$ both 
fall with density at about the same rate.

\begin{figure}[htb]
\epsscale{1.0}
\plotone{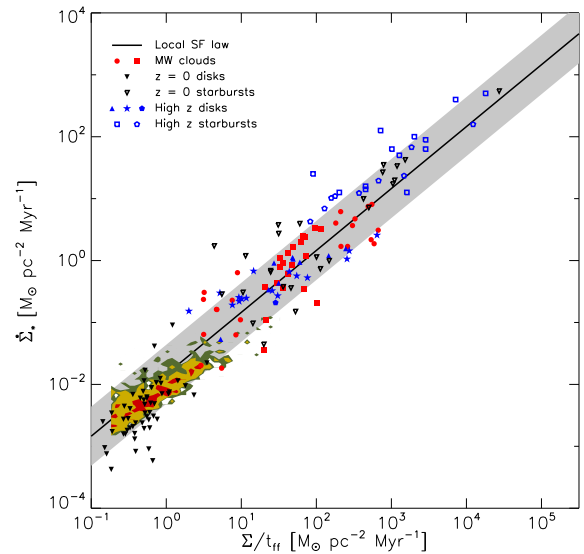}
\caption{\small
\label{fig:kdm12}
SFR per unit area versus gas column density over free-fall time \citet{Krumholz:2012}. Different shapes indicate different data sources, and colors represent different types of objects: red circles and squares are Milky Way clouds, black filled triangles and unresolved $z=0$ galaxies, black open triangles are unresolved $z=0$ starbursts, blue filled symbols are unresolved $z > 1$ disk galaxies, and blue open symbols are unresolved $z>1$ starburst galaxies. Contours show the distribution of kpc-sized regions within nearby galaxies. The black line is $\epsilon_{\rm ff} = 0.01$, and the gray band is a factor of 3 range around it. 
}
\end{figure}

Figure \ref{fig:kdm12} shows a large sample of observations compiled by \citet{Krumholz:2012}, which includes individual Galactic clouds, nearby galaxies, and high-redshift galaxies, covering a very large range of mean densities. They find that all of the data are consistent with $\epsilon_{\rm ff}\sim 0.01$, albeit with considerable scatter and systematic uncertainty. Even with the uncertainties, however, it is clear that $\epsilon_{\rm ff} \sim 1$ is strongly ruled out.

\label{sec:gmctimescales}

\bigskip
\noindent
\textbf{ 2.7. GMCs in Varying Galactic Environments} 
\bigskip

One gains useful insight into GMC physics by studying their properties as a function of environment. Some of the most extreme environments, such as those in starbursts or metal-poor galaxies, also offer unique insights into astrophysics in the primitive universe, and aid in the interpretation of observations of distant sources.

Galactic centers, which feature high metallicity and stellar density, and often high surface densities of gas and star formation, are one unusual environment to which we have observational access. The properties of the bulge, and presence of a bar appear to influence the amount of H$_2$ in the center \citep{Fisher:2013}.  
Central regions with high $\Sigma_{\rm H_2}$ preferentially show reduced $\tau_{\rm dep}({\rm H}_2)$ compared to galaxy averages \citep{Leroy:2013b}, suggesting that central GMCs convert their gas into stars more rapidly. Reduced $\tau_{\rm dep}({\rm H}_2)$ is correlated with an increase in CO (2-1)/(1-0) ratios, indicating enhanced excitation (or lower optical depth). Many galaxy centers also exhibit a super-exponential increase in CO brightness, and a drop in CO-to-H$_2$ conversion factor
\citep[][which reinforces the short $\tau_{\rm dep}({\rm H_2})$ conclusion]{Sandstrom:2012}. On the other hand, in our own Galactic Center, \citet{Longmore:2013} show that there are massive molecular clouds that have surprisingly {\em little} star formation, and depletion times $\tau_{\rm dep}({\rm H_2}) \sim 1$ Gyr comparable to disk GMCs \citep{Kruijssen:2013}, despite volume and column densities orders of magnitude higher (see Longmore et al. Chapter).

Obtaining similar spatially-resolved data on external galaxies is challenging. \citet{Rosolowsky:2005} examined several very large GMCs ($M\sim10^7$~M$_\odot$, $R\sim40-180$~pc) in M~64. They also find a size-linewidth coefficient somewhat larger than in the Milky Way disk, and, in $^{13}$CO, high surface densities.  Recent multi-wavelength, high-resolution ALMA observations of the center of the nearby starburst NGC~253 find cloud masses $M\sim10^7$~M$_\odot$ and sizes $R\sim30$~pc, implying $\Sigma_{\rm GMC}\gtrsim10^3$~M$_\odot\,{\rm pc}^{-2}$ ({\em Leroy et al. 2013, in prep.}). The cloud linewidths imply that they are self-gravitating.

The low metallicity environments of dwarf galaxies and outer galaxy disks supply another fruitful laboratory for study of the influence of environmental conditions. Because of their proximity, the Magellanic Clouds provide the best locations to study metal-poor GMCs. Owing to the scarcity of dust at low metallicity \citep[e.g.,][]{Draine:2007} the abundances of H$_2$ and CO in the ISM are greatly reduced
compared to what would be found under comparable conditions in a higher metallicity galaxy (see the discussion in \S~3.3). As a result, CO emission is faint, only being present in regions of very 
high column density \citep[e.g.,][and references therein]{Israel:1993,Bolatto:2013}. Despite these difficulties, there are a number of studies of low metallicity GMCs. \citet{Rubio:1993} reported GMCs in the SMC exhibit sizes, masses, and a size-linewidth relation similar to that in the disk of the Milky
Way. However, more recent work suggests that GMCs in the Magellanic Clouds are smaller and have
lower masses, brightness temperatures, and surface densities than typical inner Milky Way GMCs, although they are otherwise similar to Milky Way clouds \citep{Fukui:2008,Bolatto:2008,Hughes:2010,Muller:2010,Herrera:2013}. Magellanic Cloud GMCs also appear to be surrounded by extended envelopes of CO-faint H$_2$ that are $\sim30\%$ larger than the CO-emitting region \citep{Leroy:2007,Leroy:2009}. Despite their CO faintness, though, the SFR-H$_2$ relation appears to be independent of metallicity once the change in the CO-to-H$_2$ conversion factor is removed \citep{Bolatto:2011}.

\bigskip

\centerline{\textbf{3. GMC FORMATION}}
\bigskip

We now turn to the question of how GMCs form. The main mechanisms that have been proposed, which we discuss in detail below, are converging flows driven by stellar feedback or turbulence (\S~3.1), agglomeration of smaller clouds (\S~3.2), gravitational instability (\S~3.3) and magneto-gravitational instability (\S~3.4), and instability involving differential buoyancy (\S~3.5). All these mechanisms involve converging flows, i.e., $\nabla \cdot {\bf u} < 0$, where ${\bf u}$ is the velocity, which may be continuous or intermittent. Each mechanism however acts over different sizes and timescales, producing density enhancements of different magnitudes.
Consequently, different mechanisms may dominate in different environments, and may lead to different cloud properties. In addition to these physical mechanisms for gathering mass, forming a GMC involves a phase change in the ISM, and this too may happen in a way that depends on the large-scale environment (\S~3.6)

Two processes that we will not consider as cloud formation mechanisms
are thermal instabilities (TI, \citealt{Field:1965a}) and magneto-rotational instabilities (MRI, \citealt{Balbus:1991}). 
Neither of these by themselves are likely to form molecular clouds. TI produces the cold (100 K) atomic component of the ISM, but the cloudlets formed are $\sim$ pc in scale, and without the shielding provided by a large gas column this does not lead to $10-20$ K molecular gas. Furthermore, TI does not act in isolation, but interacts with all other processes taking place in the atomic ISM \citep{Vazquez-Semadeni:2000,Sanchez:2002, Piontek:2004,Kim:2008}. Nevertheless the formation of the cold atomic phase aids GMC formation by enhancing both shock compression and vertical settling, as discussed further in \S~3.2. The MRI is not fundamentally compressive, so again it will not by itself lead to GMC formation, although it can significantly affect the development of large-scale gravitational instabilities that are limited by galactic angular momentum \citep{Kim:2003}. MRI or TI may also drive turbulence \citep{Koyama:2002,Kritsuk:2002,Kim:2003,Piontek:2005,Piontek:2007,Inoue:2012}, and thereby aid cloud agglomeration and contribute to converging flows. However, except in regions with very low SFR, the amplitudes of turbulence driven by TI and MRI are lower than those driven by star formation feedback.

\bigskip
\noindent
\textbf{ 3.1 Localized Converging Flows}
\bigskip

\begin{figure}[htb]
\plotone{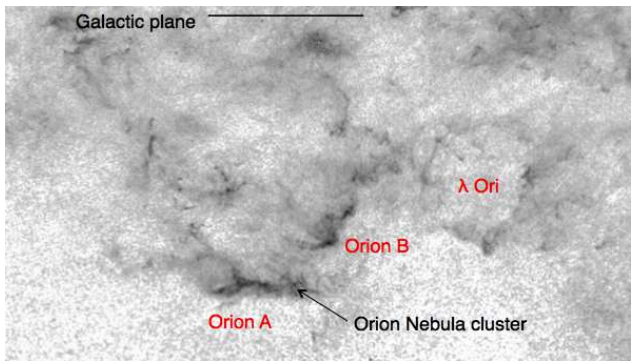}
\caption{\small
\label{fig:extinction}
Extinction map towards the Orion-Monoceros molecular complex.
Different features are located at different distances, but at
the mean distance of Orion, the approximate size of the A and B
clouds, as well as the $\lambda$ Ori ring, is of the order of 50~pc.
Note the presence of shells at a variety of scales, presumably due to
OB stars and/or supernovae. The $\lambda$ Ori ring surrounds a 5~Myr old stellar cluster, and it is thought to have been produced by a supernova
\citep[data from ][as adapted by L. Hartmann (2013, in
preparation)]{Rowles:2009}.}
\end{figure}

Stellar feedback processes such as the expansion of H~\textsc{ii} regions \citep{Bania:1980, Vazquez-Semadeni:1995, Passot:1995} and supernova blast waves \citep{McCray:1987, Gazol:1999, deAvillez:2000, deAvillez:2001, deAvillez:2005, Kim:2011, Ntormousi:2011} can drive converging streams of gas that accumulate to become molecular clouds, either in the Galactic plane or above it, or even after material ejected vertically by the local excess of pressure due to the stars/supernovae falls back into the plane of the disk. 
Morphological evidence for this process can be found in large-scale extinction maps of the Galaxy (see Fig.~\ref{fig:extinction}), and recent observations in both the Milky Way and the LMC, confirm that MCs can be found at the edges of supershells \citep{Dawson:2011,Dawson:2013}.

Locally -- on scales up to $\sim 100$ pc -- it is likely that these processes 
play a dominant role in MC formation, since on these scales the pressure due to local energy sources is typically $P/k_B \sim 10^4$ K cm$^{-3}$, which  exceeds the mean pressure of the ISM in the Solar neighborhood \citep{Draine:2011}. The mass of MCs created by this process will be defined by the mean
density $\rho_0$ and the velocity correlation length, $L$  
of the converging streams; $L$ is less than the disk scale height $H$ for local turbulence. For Solar neighborhood conditions, where there are low densities and relatively short timescales for coherent flows, 
this implies a maximum MC mass of a few times $10^4 M_\odot$. Converging flows driven by large-scale instabilities can produce higher-mass clouds (see below).

A converging flow it not by itself sufficient to form a MC; the detailed initial velocity, density, and magnetic field structure must combine with TI to produce fast cooling \citep[e.g., ][]{McCray:1975, Bania:1980, Vazquez-Semadeni:1995, Hennebelle:1999, Koyama:2000, Audit:2005, Heitsch:2006, Vazquez-Semadeni:2007}. This allows rapid accumulation of cold, dense atomic gas, and thus promotes molecule formation. The accumulation of gas preferentially along field lines (perhaps due to magneto-Jeans instability -- see \S~3.4) also increases the mass to flux ratio, causing a transition from subcritical gas to supercritical \citep{Hartmann:2001, Vazquez-Semadeni:2011}. Thus the accumulation of mass from large-scale streams, the development of a molecular phase with negligible thermal support, and the transition from magnetically subcritical to supercritical all happen essentially simultaneously, at a column density of \citep[$\sim 10^{21}$cm$^{-2}$;][see also Section~3.6]{Hartmann:2001}, allowing simultaneous molecular cloud and star formation. 
    
This mechanism for GMC formation naturally explains the small age spread observed in MCs near the Sun (see \S~2.5), since the expected star formation timescale in these models is 
the thickness of the compressed gas ($\sim$ few pc) divided by the inflow velocity ($\sim$ few km s$^{-1}$).
In addition to the overall age spread, the shape of the stellar age distribution produced by this mechanism is consistent with those observed in nearby MCs: most of the stars have ages of 1-3 Myr, and only few older. While some of these regions may be contamination by non-members \citep{Hartmann:2003}, the presence of some older stars might be the result of a few stars forming prior to the global but hierarchical and chaotic contraction concentrates most of the gas and forms most of the stars \citep{Hartmann:2012}. This model is also consistent with the observed morphology of Solar neighborhood clouds, since their elongated structures would naturally result from inflow compression.  

\begin{figure}
\plotone{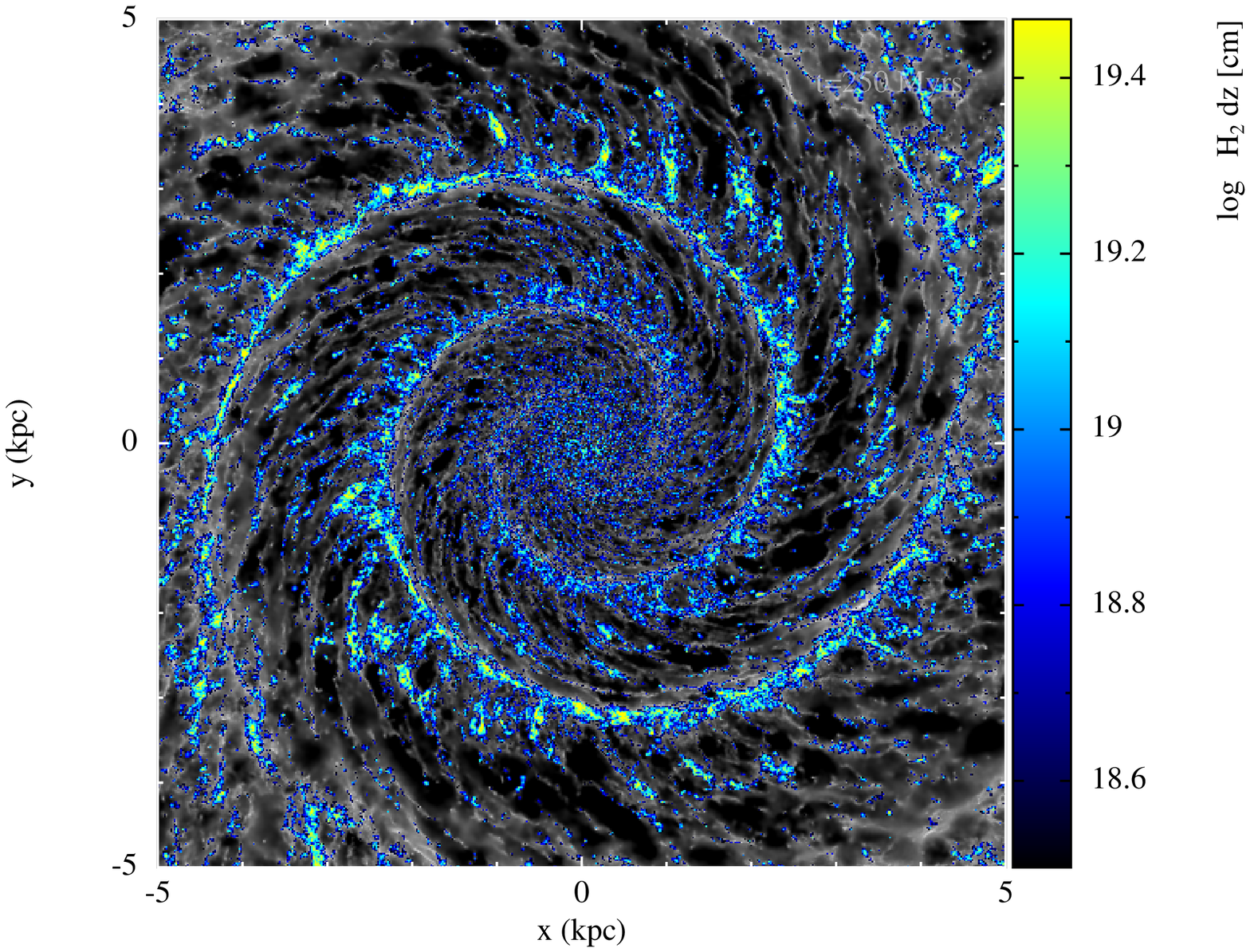}
\caption{\small
\label{fig:gmc_collisions}
A snapshot is shown from a simulation of the gas in a spiral galaxy, including a fixed sprial potential, gas heating and cooling, self gravity and stellar feedback (from \citealt{Dobbs:2013}). The GMCs in this simulation predominanlty form by cloud-cloud collisions in the spiral arms. The black and white color scheme shows the total gas column density, whilst the blue-yellow scheme shows the H$_2$ fraction integrated through the disc. }
\end{figure}

\bigskip
\noindent
\textbf{3.2 Spiral-Arm Induced Collisions}
\\
While localized converging flows can create clouds with masses up to $\sim 10^4$ $\msun$, most of the molecular gas in galaxies is found in much larger clouds (\S~2.3). 
Some mechanism is required to either form these large clouds directly, or to induce smaller clouds to agglomerate into larger ones. 
Although not yet conclusive, there is some evidence of different cloud properties in arms compared to inter-arm regions, which could suggest different mechanisms operating (Colombo et al., submitted).
It was long thought that the agglomeration of smaller clouds could not work because,
for clouds moving with observed velocity dispersions, the timescale required to build a $10^5 - 10^6$ $M_\odot$ cloud would be $>100$ Myr \citep{Blitz:1980}. However in the presence of spiral arms, collisions between clouds become much more frequent, greatly reducing the timescale \citep{Casoli:1982,Kwan:1983,Dobbs:2008a}. Fig.~\ref{fig:gmc_collisions} shows a result from a simulation where GMCs predominantly form from spiral arm induced collisions. Even in the absence of spiral arms, in high surface density galaxies \citet{Tasker:2009} find that collisions are frequent enough (every $\sim 0.2$ orbits) to influence GMC properties, and possibly star formation \citep{Tan:2000}. The frequency and success of collisions is enhanced by clouds' mutual gravity \citep{Kwan:1987,Dobbs:2008a}, but is suppressed by magnetic fields \citep{Dobbs:2008b}.

This mechanism can explain a number of observed features of GMCs. Giant molecular associations in spiral arms display a quasi-periodic spacing along the arms \citep{Elmegreen:1983,Efremov:1995}, and \citet{Dobbs:2008a} show that this spacing can be set by the epicyclic frequency imposed by the spiral perturbation, which governs the amount of material which can be collected during a spiral arm passage. A stronger spiral potential produces more massive and widely spaced clouds. The stochasticity of cloud-cloud collisions naturally produces a powerlaw GMC mass function \citep{Field:1965b,Penston:1969,Taff:1973,Handbury:1977,Kwan:1979,Hausman:1982,Tomisaka:1984}, and the powerlaw indices produced in modern hydrodynamic simulations agree well with observations \citep{Dobbs:2008a,Dobbs:2011,Tasker:2009}, provided the simulations also include a subgrid feedback recipe strong enough to prevent runaway cloud growth \citep{Dobbs:2011, Hopkins:2011}. A third signature of either local converging flows or agglomeration is that they can produce clouds that are counter-rotating compared to the galactic disk \citep{Dobbs:2008a,Dobbs:2011,Tasker:2009}, consistent with observations showing that retrograde clouds are as common as prograde ones \citep{Blitz:1993,Phillips:1999,Rosolowsky:2003,Imara:2011a,Imara:2011b}. Clouds formed by agglomeration also need not be gravitationally bound, although, again, stellar feedback is necessary to maintain an unbound population \citep{Dobbs:2011a}. 

\bigskip
\noindent
\textbf{ 3.3 Gravitational Instability}
\\

An alternative explanation for massive clouds is that they form in a direct, top-down manner, and one possible mechanism for this to happen is gravitational instability. Axisymmetric perturbations in single-phase infinitesimally thin gas disks with effective sound speed $c_{\rm eff}$, surface density $\Sigma$, and epicyclic frequency $\kappa$ can occur whenever the Toomre parameter $Q\equiv \kappa c_{\rm eff}/(\pi G \Sigma) < 1$.  For the nonaxisymmetric case, however, there are no true linear (local) instabilities because differential rotation ultimately shears any wavelet into a tightly wrapped trailing spiral with wavenumber $k\propto t$ in which pressure stabilization (contributing to the dispersion relation as $k^2 c_{\rm eff}^2$) is stronger than self-gravity (contributing as $-2 \pi G \Sigma_{\rm gas}|k|$). Using linear perturbation theory, \citet{Goldreich:1965} were the first to analyze the growth of low-amplitude shearing wavelets in gaseous disks; the analogous ``swing amplification" process for stellar disks was studied by \citet{Julian:1966}. Spiral-arm regions, because they have higher gaseous surface densities, are more susceptible than interarm regions to self-gravitating fragmentation, and this process has been analyzed using linear theory \citep[e.g.,][]{Elmegreen:1979,Balbus:1985,Balbus:1988}. Magnetic effects are particularly important in spiral-arm regions because of the reduced shear compared to interarm regions; this combination leads to a distinct process termed the magneto-Jeans instability (MJI; see \S~3.4).  

Nonaxisymmetric disturbances have higher $Q$ thresholds for growth (i.e.,~amplification is possible at lower $\Sigma$) than axisymmetric disturbances. Because the instabilities are nonlinear, numerical simulations are required to evaluate these thresholds. \citet{Kim:2001} found, for local thin-disk simulations and a range of magnetization, that the threshold is at $Q=1.2-1.4$. Vertical thickness of the disk dilutes self-gravity, which tends to reduce the critical $Q$ below unity, but magnetic fields and the contribution of stellar gravity provide compensating effects, yielding $Q_{\rm crit} \sim 1.5$ \citep{KOS:2002,Kim:2003,Kim:2007,Li:2005}.

In the absence of a pre-existing ``grand design" stellar spiral pattern, gravitational instability in the combined gas-star disk can lead to flocculent or multi-armed spirals, as modeled numerically by e.g., \citet{Li:2005,Kim:2007,Robertson:2008,Tasker:2009, Dobbs:2011,Wada:2011,Hopkins:2012}. In cases where a grand design is present (e.g., when driven by a tidal encounter), gas flowing through the spiral pattern supersonically experiences a shock, which raises the density and can trigger gravitational collapse \citep{Roberts:1969}; numerical investigations of this include e.g., \citet{Kim:2002,Kim:2006,Shetty:2006,Dobbs:2012}; see also \S~3.4 for a discussion of magnetic effects.

Clouds formed by gravitational instabilities in single-phase gas disks typically have masses $\sim 10$ times the two-dimensional Jeans mass $M_{J,2D}= c_{\rm eff}^4/(G^2 \Sigma)\sim 10^8 M_\odot (c_{\rm eff}/7\,
{\rm km\ s^{-1}})^4 (\Sigma/ M_\odot\ {\rm pc}^{-2})^{-1}$. The gathering scale for mass is larger than the 2D Jeans length $L_{J,2D} = c_{\rm eff}^2/(G \Sigma)$ in part because the fastest-growing scale exceeds $L_{J,2D}$ even for infinitesimally-thin disks, and this increases for thick disks; also, the cloud formation process is highly anisotropic because of shear.  At moderate gas surface densities, $\Sigma < 100 M_\odot \ {\rm pc}^{-2}$, as found away from spiral-arm regions, the corresponding masses are larger than those of observed GMCs when $c_{\rm eff}\sim 7\,{\rm km\ s^{-1}}$, comparable to large-scale mean velocity dispersions for the atomic medium. The absence of such massive clouds is consistent with observations indicating that $Q$ values are generally above the critical threshold (except possibly in high redshift systems), such that spiral-arm regions at high $\Sigma$ and/or processes that reduce $c_{\rm eff}$ locally are required to form observed GMCs via self-gravitating instability. While many analyses and simulations assume a single-phase medium, the diffuse ISM from which GMCs form is in fact a multi-phase medium, with cold clouds surrounded by a warmer intercloud medium. The primary contribution to the effective velocity dispersion of the cold medium is turbulence. This turbulence can dissipate due to cloud-cloud collisions, as well as large-scale flows in the horizontal direction, and flows towards the midplane from high latitude. Turbulent dissipation reduces the effective pressure support, allowing instability at lower $\Sigma$ \citep{Elmegreen:2011}. In addition, the local reduction in $c_{\rm eff}$ enables gravitational instability to form lower mass clouds. Simulations that include a multi-phase medium and/or feedback from star formation (which drives turbulence and also breaks up massive GMCs) find a broad spectrum of cloud masses, extending up to several $\times 10^6 M_{\odot}$ \citep[e.g.,][]{Wada:2000,Shetty:2008,Tasker:2009,Dobbs:2011,Hopkins:2012}.

\bigskip
\noindent
\textbf{ 3.4 Magneto-Jeans Instability}
\bigskip

In magnetized gas disks, another process, now termed the magneto-Jeans instability (MJI), can occur. This was first investigated using linear perturbation theory for disks with solid-body rotation by \citet{Lynden-Bell:1966}, and subsequently by \citet{Elmegreen:1987}, \citet{Gammie:1996}, and \citet{Kim:2001} for general rotation curves. MJI occurs in low-shear disks at any nonzero magnetization, whereas swing amplification (and its magnetized variants) described in \S~3.3 requires large shear. In MJI, magnetic tension counteracts the Coriolis forces that can otherwise suppress gravitational instability in rotating systems, and low shear is required so that self-gravity takes over before shear increases $k$ to the point of stabilization. \citet{Kim:2001} and \citet{KOS:2002} have simulated this process in uniform, low-shear regions as might be found in inner galaxies.

Because low shear is needed for MJI, it is likely to be most important either in central regions of galaxies or in spiral arms \citep{Elmegreen:1994}, where compression by a factor $\Sigma/\Sigma_0$ reduces the local shear as $d\ln \Omega/d\ln R \sim \Sigma/\Sigma_0-2$. Given the complex dynamics of spiral arm regions, this is best studied with MHD simulations \citep{Kim:2002,Kim:2006,Shetty:2006}, which show that MJI can produce massive, self-gravitating clouds either within spiral arms or downstream, depending on the strength of the spiral shock. Clouds that collapse downstream are found within overdense spurs. The spiral arms maintain their integrity better in magnetized models that in comparison unmagnetized simulations. Spacings of spiral-arm spurs formed by MJI are several times the Jeans length $c_{\rm eff}^2/(G \Sigma_{\rm arm})$ at the mean arm gas surface density $\Sigma_{\rm arm}$, consistent with observations of spurs and giant H~\textsc{ii} regions that form ``beads on a string" in grand design spirals \citep{Elmegreen:1983,Lavigne:2006}. The masses of gas clumps formed by MJI are typically $\sim 10^7 M_\odot$ in simulations with $c_{\rm eff} = 7$ km s$^{-1}$, comparable to the most massive observed giant molecular cloud complexes and consistent with expectations from linear theory. Simulations of MJI in multiphase disks with feedback-driven turbulence have not yet been conducted, so there are no predictions for the full cloud spectrum.

\bigskip
\noindent
\textbf{ 3.5 Parker Instability}
\\

A final possible top-down formation mechanism is Parker instability, in which horizontal magnetic fields threading the ISM buckle due to differential buoyancy in the gravitational field, producing dense gas accumulations at the midplane. Because the most unstable modes have wavelengths $\sim (1-2)2 \pi H$, where $H$ is the disk thickness, gas could in principle be gathered from scales of several hundred pc to produce quite massive clouds \citep{Mouschovias:1974,Mouschovias:1974b,Blitz:1980}. However, Parker instabilities are self-limiting \citep[e.g.,][]{Matsumoto:1988}, and in single-phase media saturate at factor of few density enhancements at the midplane \citep[e.g.,][]{Basu:1997,Santillan:2000,Machida:2009}. Simulations have also demonstrated that three-dimensional dynamics, which enhance reconnection and vertical magnetic flux redistribution, result in end states with relatively uniform horizontal density distributions \citep[e.g.,][]{Kim:1998,Kim:2003}. Some models suggest that spiral arm regions may be favorable for undular modes to dominate over interchange ones \citep{Franco:2002}, but simulations including self-consistent flow through the spiral arm indicate that vertical gradients in horizontal velocity (a consequence of vertically-curved spiral shocks) limit the development of the undular Parker mode \citep{Kim:2006}.  While Parker instability has important consequences for vertical redistribution of magnetic flux, and there is strong evidence for the formation of magnetic loops anchored by GMCs in regions of high magnetic field strength such as the Galactic Center \citep{Fukui:2006,Torii:2010}, it is less clear if Parker instability can create massive, highly overdense clouds.

For a medium subject to TI, cooling of overdense gas in magnetic valleys can strongly enhance the density contrast of structures that grow by Parker instability \citep{Kosinski:2007,Mouschovias:2009}. However, simulations to date have not considered the more realistic case of a pre-existing cloud / intercloud medium in which turbulence is also present. For a non-turbulent medium, cold clouds could easily slide into magnetic valleys, but large turbulent velocities may prevent this. In addition, whatever mechanisms drive turbulence may disrupt the coherent development of large-scale Parker modes.  An important task for future modeling of the multiphase, turbulent ISM is to determine whether Parker instability primarily re-distributes the magnetic field and alters the distribution of low-density coronal gas, or if it also plays a major role in creating massive, bound GMCs near the midplane.

\bigskip
\noindent
\textbf{ 3.6 Conversion of H to H$_2$, and C$^+$ to CO}
\\

Thus far our discussion of GMC formation has focused on the mechanisms for accumulating high density gas. However, the actual observable that defines a GMC is usually CO emission, or in some limited cases other tracers of H$_2$ \citep[e.g.,][]{Bolatto:2011}. Thus we must also consider the chemical transition from atomic gas, where the hydrogen is mostly H~\textsc{i} and carbon is mostly C$^+$, to molecular gas characterized by H$_2$ and CO. The region over which this transition occurs is called a photodissociation or photon-dominated region (PDR). H$_2$ molecules form (primarily) on the surfaces of dust grains, and are destroyed by resonant absorption of Lyman- and Werner-band photons. The equilibrium H$_2$ abundance is controlled by the ratio of the far ultraviolet (FUV) radiation field to the gas density, so that H$_2$ becomes dominant in regions where the gas is dense and the FUV is attenuated \citep{vanDishoeck:1986,Black:1987,Sternberg:1988,Krumholz:2008,Krumholz:2009c,Wolfire:2010}. Once H$_2$ is abundant, it can serve as the seed for fast ion-neutral reactions that ultimately culminate in the formation of CO. Nonetheless CO is also subject to photodissociation, and it requires even more shielding than H$_2$ before it becomes the dominant C repository \citep{vanDishoeck:1988}.

The need for high extinction to form H$_2$ and CO has two important consequences. One, already alluded to in \S~2, is that the transition between H~\textsc{i} and H$_2$ is shifted to higher surface densities in galaxies with low metallicity and dust abundance and thus low extinction per unit mass \citep{Fumagalli:2010, Bolatto:2011, Wong:2013}. A second is that the conversion factor $X_{\rm CO}$ between CO emission and H$_2$ mass increases significantly in low metallicity galaxies \citep[and references therein]{Bolatto:2013}, because the shift in column density is much greater for CO than for H$_2$. CO and H$_2$ behave differently because self-shielding against dissociation is much stronger for H$_2$ than for CO, which must rely on dust shielding \citep{Wolfire:2010}. As a result, low metallicity galaxies show significant CO emission only from high-extinction peaks of the H$_2$ distribution, rather than from the bulk of the molecular material.

While the chemistry of GMC formation is unquestionably important for understanding and interpreting observations, the connection between chemistry and dynamics is considerably less clear. In part this is due to numerical limitations. At the resolutions achievable in galactic-scale simulations, one can model H$_2$ formation only via subgrid models that are either purely analytic \citep[e.g.,][]{Krumholz:2008, Krumholz:2009c, Krumholz:2009b, McKee:2010, Narayanan:2011, Narayanan:2012, Kuhlen:2012, Jaacks:2013} or that solve the chemical rate equations with added ``clumping factors" that are tuned to reproduce observations \citep[e.g.,][]{Robertson:2008, Gnedin:2009, Pelupessy:2009, Christensen:2012}. Fully-time-dependent chemodynamical simulations that do not rely on such methods are restricted to either simple low-dimensional geometries \citep{Koyama:2000,Bergin:2004} or to regions smaller than typical GMCs \citep{Glover:2007a, Glover:2007b, Glover:2010, Glover:2011, Clark:2012, Inoue:2012}. 

One active area of research is whether the H~\textsc{i} to H$_2$ or C$^+$ to CO transitions are either necessary or sufficient for star formation. For CO the answer appears to be ``neither". The nearly fixed ratio of CO surface brightness to star formation rate we measure in Solar metallicity regions (see \S~2.1) drops dramatically in low metallicity ones \citep{Gardan:2007, Wyder:2009, Bolatto:2011, Bolatto:2013}, strongly suggesting that the loss of metals is changing the carbon chemistry but not the way stars form. Numerical simulations suggest that, at Solar metallicity, CO formation is so rapid that even a cloud undergoing free-fall collapse will be CO-emitting by the time there is substantial star formation \citep{Hartmann:2001, Heitsch:2008}. Conversely, CO can form in shocks even if the gas is not self-gravitating \citep{Dobbs:2008c, Inoue:2012}. Moreover, formation of CO does not strongly affect the temperature of molecular clouds, so it does not contribute to the loss of thermal support and onset of collapse \citep{Krumholz:2011, Glover:2012}. Thus it appears that CO accompanies star formation, but is not causally related to it.

The situation for H$_2$ is much less clear. Unlike CO, the correlation between star formation-H$_2$ correlation appears to be metallicity-independent, and is always stronger than the star formation-total gas correlation. Reducing the metallicity of a galaxy at fixed gas surface density lower both the H$_2$ abundance and star formation rate, and by nearly the same factor \citep{Wolfe:2006, Rafelski:2011, Bolatto:2011}. This still does not prove causality, however. \citet{Krumholz:2011} and \citet{Glover:2012} suggest that the explanation is that both H$_2$ formation and star formation are triggered by shielding effects. Only at large extinction is the photodissociation rate suppressed enough to allow H$_2$ to become dominant, but the same photons that dissociate H$_2$ also heat the gas via the grain photoelectric effect, and gas only gets cold enough to form stars where the photoelectric effect is suppressed. This is, however, only one possible explanation of the data.

A final question is whether the chemical conditions in GMCs are in equilibrium or non-equilibrium. The gas-phase ion-neutral 
reactions that lead to CO formation have high rates even at low temperatures, so carbon chemistry is likely to be in equilibrium \citep{Glover:2010, Glover:2011, Glover:2012b}. The situation for H$_2$ is less clear. The rate coefficient for H$_2$ formation on dust grain surfaces is quite low, so whether gas can reach equilibrium depends on the density structure and the time available. Averaged over $\sim 100$ pc size scales and at metallicities above $\sim 1\%$ of Solar, \citet{Krumholz:2011a} find that equilibrium models of \citet{Krumholz:2009c} agree very well with time-dependent ones by \citet{Gnedin:2009}, and the observed metallicity-dependence of the H~\textsc{i}-H$_2$ transition in external galaxies is also consistent with the predictions of the equilibrium models \citep{Fumagalli:2010, Bolatto:2011, Wong:2013}. However, \citet{Mac-Low:2012} find that equilibrium models do not reproduce their simulations on $\sim 1-10$ pc scales. Nonetheless, observations of Solar Neighborhood clouds on such scales appear to be consistent with equilbrium \citep{Lee:2012}. All models agree, however, that non-equilibrium effects must become dominant at metallicities below $\sim 1-10\%$ of Solar, due to the reduction in the rate coefficient for H$_2$ formation that accompanies the loss of dust \citep{Krumholz:2012c, Glover:2012b}.

\bigskip
\bigskip
\centerline{\textbf{4. STRUCTURE, EVOLUTION, AND DESTRUCTION}}
\bigskip

Now that we have sketched out how GMCs come into existence, we consider the processes that drive their internal structure, evolution, and eventual dispersal.

\bigskip
\noindent
\textbf{4.1 GMC Internal Structure}
\bigskip

GMCs are characterized by a very clumpy, filamentary structure (see the reviews by \textit{Andr\'e et al.} and \textit{Molinari et al.} in this volume), which can be produced by a wide range of processes, including gravitational collapse \citep[e.g.,][] {Larson:1985, Nagai:1998, Curry:2000, Burkert:2004}, non-self-gravitating supersonic turbulence \citep{Padoan:2001}, and colliding flows plus thermal instability \citep{Audit:2005, Vazquez-Semadeni:2006}; all these processes can be magnetized or unmagnetized. Some recent observational studies argue that the morphology is consistent with multi-scale infall \citep[e.g.,][]{Galvan:2009, Schneider:2010, Kirk:2013}, but strong conclusions will require quantitative comparison with a wide range of simulations.

One promising approach to such comparisons is to develop statistical measures that can be applied to both simulations and observations, either two-dimensional column density maps or three-dimensional position-position-velocity cubes. \citet{Padoan:2004, Padoan:2004b} provide one example. They compare column density PDFs produced in simulations of sub-Alfv\'enic and super-Alfv\'enic turbulence, and argue that observed PDFs are better fit by the super-Alfv\'enic model. This is in some tension with observations showing that magnetic fields remain well-ordered over a wide range of length scales (see the recent review by \citealt{Crutcher:2012} and Li et al., this volume). 
The need for super-Alfv\'enic turbulence in the simulations may arise from the fact that they did not include self-gravity, thus requiring stronger turbulence to match the observed level of structure \citep{Vazquez-Semadeni:2008}. Nevertheless, the \citeauthor{Padoan:2004} results probably do show that magnetic fields cannot be strong enough to render GMCs sub-Alfv\'enic.

A second example comes from \citet{Brunt:2010a}, \citet{Brunt:2010c, Brunt:2010b} and \citet{Price:2011}, who use the statistics of the observed 2D column density PDF to infer the underlying 3D volume density PDF, and in turn use this to constrain the relationship between density variance and Mach number in nearby molecular clouds. They conclude from this analysis that a significant fraction of the energy injection that produces turbulence must be in compressive rather than solenoidal modes. Various authors \citep{Kainulainen:2009, Kritsuk:2011, Ballesteros:2011, Federrath:2013} also argue that the statistics of the density field are also highly sensitive to the amount of star formation that has taken place in a cloud, and can therefore be used as a measure of evolutionary state.

\bigskip
\noindent
\textbf{4.2 Origin of Nonthermal Motions}
\bigskip

As discussed in \S~2.3, GMCs contain strong nonthermal motions, with the bulk of the energy in modes with size scales comparable to the cloud as a whole. For a typical GMC density of $\sim 100$ cm$^{-3}$, temperature of $\sim 10$ K, and bulk velocity of $\sim 1$ km s$^{-1}$, the viscous dissipation scale is $\sim 10^{12}$ cm ($\sim$0.1 AU), implying that the Reynolds number of these motions is $\sim 10^9$.
Such a high value of the Reynolds number essentially guarantees that the flow will be turbulent.
Moreover, since the bulk velocity greatly exceeds the sound speed, the turbulence must be supersonic, though not necessarily super-Alfv\'{e}nic.
\citet{Zuckerman:1974b} proposed that this turbulence would be
confined to small scales, but modern simulations of supersonic turbulence indicate that the power is mostly on large scales. It is also possible that the linewidths contain a significant contribution from coherent infall, as we discuss below.

While turbulence is 
expected, the deeper question is why the linewidths are so large in the first place. Simulations conducted over the last $\sim 15$ years have generally demonstrated that, in the absence of external energy input, turbulence decays in $\sim 1$ crossing time of the outer scale of the turbulent flow \citep{Mac-low:1998, Mac-low:1999, Stone:1998, Padoan:1999}, except in the case of imbalanced MHD turbulence \citep{Cho:2002}. Thus the large linewidths observed in GMCs would not in general be maintained for long periods in the absence of some external input. This problem has given rise to a number of proposed solutions, which can be broadly divided into three categories: global collapse, externally-driven turbulence, and internally-driven turbulence.

\bigskip
\noindent
\textbf{4.2.1 The Global Collapse Scenario}
\bigskip

The global collapse scenario, first proposed by \citet{Goldreich:1974} and \citet{Liszt:1974},
and more recently revived by \citet{Vazquez-Semadeni:2007, Vazquez-Semadeni:2009}, \citet{Heitsch:2008,Heitsch:2008b}, \citet{Heitsch:2009}, \citet{Ballesteros:2011a, Ballesteros:2011}, and \citet{Hartmann:2012} as a nonlinear version of the hierarchical fragmentation scenario proposed by \citet{Hoyle:1953}, offers perhaps the simplest solution: the linewidths are dominated by global gravitational collapse rather than random turbulence. This both provides a natural energy source (gravity) and removes the need to explain why the linewidths do not decay, because in this scenario GMCs, filaments, and clumps are not objects 
that need to be supported, but rather constitute a hierarchy of stages in a global, highly inhomogeneous collapse flow, with each stage accreting from its parent \citep{Vazquez-Semadeni:2009}.

Investigations of this scenario generally begin by considering an idealized head-on collision between two single-phase, warm, diffuse gas streams, which might be caused by either local feedback or large-scale gravitational instability (cf.\ \S~3). (Simulations of more realistic glancing collisions between streams already containing dense clumps have yet to be performed.) The large scale compression triggers the formation of a cold cloud, which quickly acquires many Jeans masses because the warm-cold phase transition causes a factor of $\sim 100$ increase in density and
a decrease by the same factor in temparature.
Thus the Jeans mass, $M_{\rm J} \propto \rho^{-1/2} T^{3/2}$, decreases by a factor $\sim 10^4$ \citep[e.g.,][]{Vazquez-Semadeni:2012}. The cloud therefore readily fragments into clumps \citep{Heitsch:2005, Vazquez-Semadeni:2006}, and the ensemble of clumps becomes gravitationally unstable and begins an essentially pressure-free collapse. It contracts first along its shortest dimension \citep{Lin:1965}, producing sheets and then filaments \citep{Burkert:2004, Hartmann:2007, Vazquez-Semadeni:2007, Vazquez-Semadeni:2010, Vazquez-Semadeni:2011, Heitsch:2008, Heitsch:2009}. Although initially the motions in the clouds are random and turbulent, they become ever-more infall-dominated as the collapse proceeds. However, because these motions have a gravitational origin, they naturally appear virialized \citep{Vazquez-Semadeni:2007, Ballesteros:2011a}. Accretion flows consistent with the scenario have been reported in several observational studies of dense molecular gas \citep[e.g.,][]{Galvan:2009, Schneider:2010, Kirk:2013}, but observations have yet to detect the predicted inflows at the early, large-scale stages of the hierarchy. These are difficult to detect because it is not easy to separate the atomic medium directly connected to molecular clouds from the general H~\textsc{i} in the galaxy, and because the GMCs are highly fragmented, blurring the inverse p-Cygni profiles expected for infall. 
In fact,  \citet{Heitsch:2009} show that the CO line profiles of chaotically collapsing clouds match observations of GMCs. 

While this idea elegantly resolves the problem of the large linewidths, it faces challenges with respect to the constraints imposed by the observed $20-30$ Myr lifetimes of GMCs (\S~2.5) and the low rates and efficiencies of star formation (\S~2.6). We defer the question of star formation rates and efficiencies to \S~5. Concerning GMC lifetimes, a semi-analytic model by \citet{Zamora:2012} for the evolution of the cloud mass and SFR in this scenario shows agreement with the observations within factors of a few. Slightly smaller timescales ($\sim 10$ Myr) are observed in numerical simulations of cloud build-up that consider the evolution of the molecular content of the cloud \citep{Heitsch:2008}, although these authors considered substantially smaller cloud masses 
and more dense flows. 
Simulations considering larger cloud masses (several $10^4 \Msun$) exhibit evolutionary timescales $\sim 20$ Myr \citep{Colin:2013}, albeit no tracking of the molecular fraction was performed there. Simulations based on the global collapse scenario have not yet examined the formation and evolution of clouds of masses above $10^5 \Msun$, comparable to those studied in extragalactic observations. Moreover, in all of these models the lifetime depends critically on the duration and properties of the gas inflow that forms the clouds, which is imposed as a boundary condition. Self-consistent simulations in which the required inflows are generated by galactic scale flows also remain to be performed.

\bigskip
\noindent
\textbf{4.2.2 The External Driving Scenario}
\bigskip

The alternative possibility is that the large linewidths of GMCs are dominated by random motions rather than global collapse. This would naturally explain relatively long GMC lifetimes and (as we discuss in \S~5) low star formation rates, but in turn raises the problem of why these motions do not decay, giving rise to a global collapse. The external driving scenario proposes that this decay is offset by the injection of energy by flows external to the GMC. One obvious source of such external energy is the accretion flow from which the cloud itself forms, which can be subject to non-linear thin shell instability \citep{Vishniac:1994} or oscillatory overstability \citep{Chevalier:1982} that will drive turbulence. \citet{Klessen:2010} point out that only a small fraction of the gravitational potential energy of material accreting onto a GMC would need to be converted to bulk motion before it is dissipated in shocks in order to explain the observed linewidths of GMCs, and semi-analytic models by \citet{Goldbaum:2011} confirm this conclusion. Numerical simulations confirm that cold dense layers confined by the ram pressure of accretion flows indeed are often turbulent \citep{Hunter:1986, Stevens:1992, Walder:2000, Koyama:2002, Audit:2005, Heitsch:2005, Vazquez-Semadeni:2006}, although numerical simulations consistently show that the velocity dispersions of these flows are significantly smaller than those observed in GMCs unless the flows are self-gravitating \citep{Koyama:2002, Heitsch:2005, Vazquez-Semadeni:2007, Vazquez-Semadeni:2010}. This can be understood because the condition of simultaneous thermal and ram pressure balance implies that the Mach numbers in both the warm and cold phases are comparable \citep{Banerjee:2009}.

While accretion flows are one possible source of energy, there are also others. Galactic-scale and kpc-scale simulations by \citet{Tasker:2009}, \citet{Tasker:2011}, \citet{Dobbs:2011, Dobbs:2011a, Dobbs:2012}, \citet{Dobbs:2013}, \citet{Hopkins:2012b}, \citet{Van-Loo:2013} all show that GMCs are embedded in large-scale galactic flows that subject them to continuous external buffeting -- cloud-cloud collisions, encounters with shear near spiral arms, etc. -- even when the cloud's mass is not necessarily growing. These external motions are particularly important for the most massive clouds, which preferentialy form via large-scale galactic flows, and can drive turbulence in them over a time significantly longer than $\tau_{\rm ff}$. This mechanism seems particularly likely to operate in high-surface density galaxies where the entire ISM is molecular and thus there is no real distinction between GMCs and other gas, and in fact seems to be required to explain the large velocity dispersions observed in high-redshift galaxies \citep{Krumholz:2010}.

\bigskip
\noindent
\textbf{4.2.3 The Internal Driving Scenario}
\bigskip

The internally-driven scenario proposes that stellar feedback internal to a molecular cloud is responsible for driving turbulence and explaining the large linewidths seen,
in conjunction with externally-driven turbulence in the very rare clouds without significant star formation \citep[e.g., the so-called Maddalena's cloud;][]{Williams:1994}. There are a number of possible sources of turbulent driving, including H~\textsc{ii} regions, radiation pressure, protostellar outflows, and the winds of main sequence stars. (Supernovae are unlikely to be important as
an internal driver of turbulence in GMCs in most galaxies because the stellar evolution timescale is comparable to the crossing timescale, 
though they could be important as an external driver, for the dispersal of GMCs, and in starburst galaxies, see below). \citet{Matzner:2002} and \citet{Dekel:2013} provide useful summaries of the momentum budgets associated with each of these mechanisms, and they are discussed in much more detail in the chapter by \textit{Krumholz et al.}~in this volume.

H~\textsc{ii} regions are one possible turbulent driver. \citet{Krumholz:2006} and \citet{Goldbaum:2011} conclude from their semi-analytic models that H~\textsc{ii} regions provide a power source sufficient to offset the decay of turbulence, that most of this power is distributed into size scales comparable to the size of the cloud, and that feedback power is comparable to accretion power. 
The results from simulations are less clear. \citet{Gritschneder:2009} and \citet{Walch:2012} find that H~\textsc{ii} regions in their simulations drive turbulence at velocity dispersions comparable to observed values, while \citet{Dale:2005, Dale:2012, Dale:2013} and \citet{Colin:2013} find that H~\textsc{ii} regions rapidly disrupt GMCs with masses up to $\sim 10^5 \Msun$ within less than 10 Myr (consistent with observations showing that $>10$ Myr-old star clusters are usually gas-free -- \citealt{Leisawitz:1989, Mayya:2012}), but do not drive turbulence. The origin of the difference is not clear, as the simulations differ in several ways, including the geometry they assume, the size scales they consider, and the way that they set up the initial conditions.

\begin{figure}
\plotone{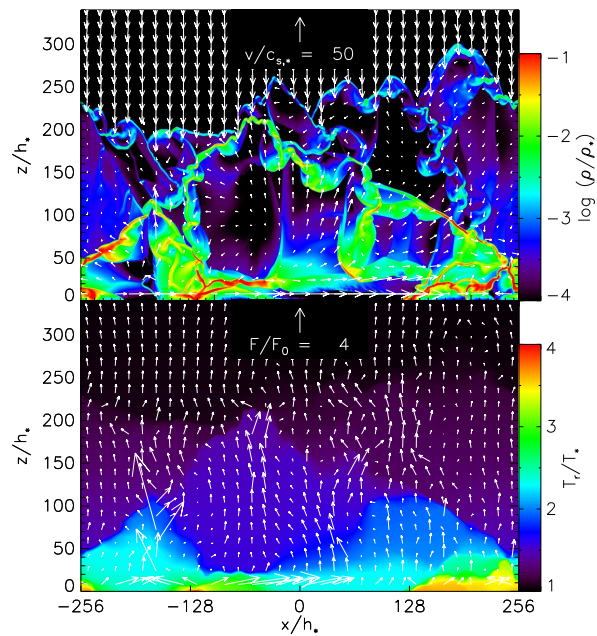}
\caption{\small
\label{fig:RRTI}
Time slice from a radiation-hydrodyanmic simulation of a molecular cloud with a strong radiation flux passing through it. In the top panel, color shows density and arrows show velocity; in the bottom panel, color shows temperature and arrows show radiation flux. Density and temperature are both normalized to characteristic values at the cloud edge; velocity is normalized to the sound speed at the cloud edge, and flux to the injected radiation flux. The simulation demonstrates that radiation pressure-driven turbulence is possible, but also that the required radiation flux and matter column density are vastly in excess of the values found in real GMCs. Figure taken from \citet{krumholz:2012b}.
}
\end{figure}

Nevertheless,
in GMCs where the escape speed approaches the 10 km s$^{-1}$ sound speed in photoionized gas, H~\textsc{ii} regions can no longer drive turbulence nor disrupt the clouds, 
and some authors have proposed that radiation pressure might take over \citep{thompson:2005, krumholz:2009, fall:2010, murray:2010, Hopkins:2011, Hopkins:2012b}.  Simulations on this point are far more limited, and the only ones published so far that actually include radiative transfer (as opposed to a sub-grid model for radiation pressure feedback) are those of \citet{krumholz:2012b, krumholz:2013}, who conclude that radiation pressure is unlikely to be important on the scales of GMCs. Figure \ref{fig:RRTI} shows a result from one of these simulations.

Stellar winds and outflows can also drive turbulence. Winds from hot main-sequence stars have been studied by \citet{Dale:2008} and \citet{Rogers:2013}, who find that they tend to ablate the clouds but not to drive significant turbulence. On the other hand, studies of protostellar outflows find that on small scales they can maintain a constant velocity dispersion while simultaneously keeping the star formation rate low \citep{Cunningham:2006, Li:2006, Matzner:2007, Nakamura:2007, Wang:2010}. On the other hand, these studies also indicate that outflows cannot be the dominant driver of turbulence on $\sim 10-100$ pc scales in GMCs, both because they lack sufficient power, and because they tend to produce a turbulent power spectrum with a distinct bump at $\sim 1$ pc scales, in contrast to the pure powerlaw usually observed.

Whether any of these mechanisms can be the dominant source of the large linewidths seen in GMCs remains unsettled. One important caveat is that only a few of the simulations with feedback have included magnetic fields, and \citet{Wang:2010} and \citet{Gendelev:2012} show (for protostellar outflows and H~\textsc{ii} regions, respectively) that magnetic fields can dramatically increase the ability of internal mechanisms to drive turbulence, because they provide an effective means of transmitting momentum into otherwise difficult-to-reach portions of clouds. MHD simulations of feedback in GMCs are clearly needed.

\bigskip
\noindent
\textbf{ 4.3 Mass Loss and Disruption}
\bigskip

As discussed in \S~2.5, GMCs are disrupted long before they can turn a significant fraction of their mass into stars. The question of what mechanism or mechanisms are responsible for this is closely tied to the question of the origin of GMC turbulence discussed in the previous section, as each proposed answer to that question imposes certain requirements on how GMCs must disrupt. In the global collapse scenario, disruption must occur in less than the mean-density free-fall time to avoid excessive star formation.  
Recent results suggest a somewhat slower collapse in flattened or filamentary objects \citep{Toala:2012, Pon:2012}, but disruption must still be fast. In the externally-driven or internally-driven turbulence scenarios disruption can be slower, but must still occur before the bulk of the material can be converted to stars. Radiation pressure and protostellar outflows appear unlikely to be responsible, for the same reasons (discussed in the previous section) that they cannot drive GMC-scale turbulence. For main sequence winds, \citet{Dale:2008} and \citet{Rogers:2013} find that they can expel mass from small, dense molecular clumps, but it is unclear if the same is true of the much larger masses and lower density scales that characterize GMCs.

The remaining stellar feedback mechanisms, photoionization and supernovae, are more promising. Analytic models have long predicted that photoionization should be the primary mechanism for ablating mass from GMCs \citep[e.g.,][]{Field:1970, Whitworth:1979, Cox:1983, Williams:1997, Matzner:2002, Krumholz:2006, Goldbaum:2011}, and numerical simulations by \citet{Dale:2012,Dale:2013} and \citet{Colin:2013} confirm that photoionization is able to disrupt GMCs with masses of $\sim 10^5$ $M_\odot$ or less. More massive clouds, however, may have escape speeds too high for photoionization to disrupt them unless they suffer significant ablation first.

Supernovae are potentially effective in clouds of all masses, but are in need of further study. Of cataloged Galactic supernova remnants, 8\% (and 25\% of X-ray emitting remnants) are classified as ``mixed morphology,'' believed to indicate interaction between the remnant and dense molecular gas \citep{Rho:1998}. This suggests that a non-negligible fraction of GMCs may interact with supernovae. Because GMCs are clumpy, this interaction will differ from the standard solutions in a uniform medium
\citep[e.g.,][]{Cioffi:1988,Blondin:1998}, but theoretical studies of supernova remnants in molecular gas have thus far focused mainly on emission properties \citep[e.g.,][]{Chevalier:1999,Chevalier:2001,Tilley:2006} rather than the dynamical consequences for GMC evolution. Although some preliminary work \citep{Kovalenko:2012} suggests that an outer shell will still form, with internal clumps accelerated outward when they are overrun by the expanding shock front \citep[cf.][]{Klein:1994,Mac-Low:1994}, complete simulations of supernova remant expansion within realistic GMC enviroments are lacking.  Obtaining a quantitative assessment of the kinetic energy and momentum imparted to the dense gas will be crucial for understanding  GMC destruction.

\bigskip

\centerline{\textbf{ 5. Regulation of star formation in GMCs}}
\bigskip

Our discussion thus far provides the framework to address the final topic of this review: what are the dominant interstellar processes that regulate the rate of star formation at GMC and galactic scales? The accumulation of GMCs is the first step in star formation, and large scale, top-down processes appear to determine a cloud's starting mean density, mass to magnetic flux ratio, Mach number, and boundedness. But are these initial conditions retained for times longer than a cloud dynamical time, and do they affect the formation of stars within the cloud? If so, how stars form is ultimately determined by the large scale dynamics of the host galaxy. Alternatively, if the initial state of GMCs is quickly erased by internally-driven turbulence or external perturbations, then the regulatory agent of star formation lies instead on small scales within individual GMCs. In this section, we review the proposed schemes and key GMC properties that regulate the production of stars.

\bigskip
\noindent
\textbf{ 5.1 Regulation Mechanisms}
\bigskip

Star formation occurs at a much lower pace than its theoretical possible 
free-fall maximum (see Section 2.6). 
Explaining why this is so is a key goal of star formation theories. These theories are intimately related to the assumptions made about the evolutionary path of GMCs. Two theoretical limits for cloud evolution are a state of global collapse with a duration $\sim \tau_{\rm ff}$ and a quasi-steady state in which 
clouds are supported for times $\gg \tau_{\rm ff}$. 

In the global collapse limit, one achieves low SFRs by having a low net star formation efficiency $\epsilon_*$ over the lifetime $\sim \tau_{\rm ff}$ of any given GMC, and then disrupting the GMC via feedback. The mechanisms invoked to accomplish this are the same as those invoked in Section 4.2.3 to drive internal turbulence: photoionization and supernovae. Some simulations suggest this these mechanisms can indeed enforce low $\epsilon_*$: \citet{Vazquez-Semadeni:2010} and \citet{Colin:2013}, using a subgrid model for ionizing feedback, find that $\epsilon_* \lesssim 10\%$ for clouds up to $\sim 10^5$ $M_\odot$, and \citet{Zamora:2012} find
that the evolutionary timescales produced by this mechanism of cloud disruption are consistent with those inferred in the Large Magellanic Cloud (Section 2.5). On the other hand, 
it remains unclear
what mechanisms might be able to disrupt $\sim 10^6$ $M_\odot$ clouds.

If clouds are supported against large-scale collapse, then star formation consists of a small fraction of the mass ``percolating'' through this support to collapse and form stars. Two major forms of support have been considered: magnetic \citep[e.g.,][]{Shu:1987, Mouschovias:1991a, Mouschovias:1991b} and turbulent \citep[e.g.,][]{Mac-Low:2004, Ballesteros:2007}. While dominant for over two decades, the magnetic support theories, in which the percolation was allowed by ambipolar diffusion, are now less favored, (though see \citealt{Mouschovias:2009, Mouschovias:2010}) due to growing observational evidence that molecular clouds are magnetically supercritical (Section 2.4). We do not discuss these models further. 

In the turbulent support scenario, supersonic turbulent motions behave as a source of pressure with respect to structures whose size scales are larger than the largest scales of the turbulent motions (the ``energy containing scale'' of the turbulence), while inducing local compressions at scales much smaller than that. A simple analytic argument suggests that, regardless of whether turbulence is internally- or externally-driven, its net effect is to increase the effective Jeans mass as $M_{\rm J,turb} \propto v_{\rm rms}^2$, where $v_{\rm rms}$ is the rms turbulent velocity \citep{Mac-Low:2004}. Early numerical simulations of driven turbulence in isothermal clouds \citep{Klessen:2000, vazquez-semadeni:2003} indeed show that, holding all other quantities fixed,
raising the Mach number of the flow decreases the dimensionless star formation rate $\epsilon_{\rm ff}$. However, this is true only as long as the turbulence is maintained; if it is allowed to decay, then raising the Mach number actually raises $\epsilon_{\rm ff}$, because in this case the turbulence simply accelerates the formation of dense regions and then dissipates \citep{Nakamura:2005}.
Magnetic fields, even those not strong enough to render the gas subcritical, also decrease $\epsilon_{\rm ff}$ \citep{Heitsch:2001, Vazquez-Semadeni:2005, Padoan:2011, Federrath:2012}.

To calculate the SFR in this scenario, one can idealize the turbulence level, mean cloud density, and SFR as quasi-stationary, and then attempt to compute $\epsilon_{\rm ff}$. In recent years, a number of analytic models have been developed to do so (\citealt{Krumholz:2005, Padoan:2011, Hennebelle:2011}; see \citealt{Federrath:2012} for a useful compilation, and for generalizations of several of the models). These models generally exploit the fact that supersonic isothermal turbulence produces a probability density distribution (PDF) with a lognormal form \citep{Vazquez-Semadeni:1994}, so that there is always a fraction of the mass at high densities. The models then assume that the mass at high densities (above some threshold), $M_{\rm hd}$, is responsible for the instantaneous SFR, which is given as SFR=$M_{\rm hd}/\tau$, where $\tau$ is some characteristic timescale of the collapse at those high densities.

In all of these models $\epsilon_{\rm ff}$ is determined by other dimensionelss numbers: the rms turbulent Mach number ${\cal M}$, the virial ratio $\alpha_G$, and (when magnetic fields are considered) the magnetic $\beta$ parameter; the ratio of compressive to solenoidal modes in the turbulence is a fourth possible parameter \citep{Federrath:2008, Federrath:2012}. The models differ in their choices of density threshold and timescale (see the chapter by \textit{Padoan et al.}), leading to variations in the predicted dependence of $\epsilon_{\rm ff}$ on $\mathcal{M}$, $\alpha_G$, and $\beta$. However, all the models produce $\epsilon_{\rm ff} \sim 0.01 - 0.1$ for dimensionless values comparable to those observed. \citet{Federrath:2012} and \citet{Padoan:2012} have conducted large campaigns of numerical simulations where they have systematically varied $\mathcal{M}$, $\alpha_G$, and $\beta$, measured $\epsilon_{\rm ff}$, and compared to the analytic models. \citet{Padoan:2012} give their results in terms of the ratio $t_{\rm ff}/t_{\rm dyn}$ rather than $\alpha_G$, but the two are identical up to a constant factor \citep{Tan:2006}. In general they find that $\epsilon_{\rm ff}$ decreases strongly with $\alpha_G$ and increases weakly with $\mathcal{M}$, and that a dynamically-significant magnetic field (but not one so strong as to render the gas subcritical) reduces $\epsilon_{\rm ff}$ by a factor of $\sim 3$. Simulations produce $\epsilon_{\rm ff} \sim 0.01 - 0.1$, in general agreement with the range of analytic predictions. 

\begin{figure}[htb]
\epsscale{1.0}
\plotone{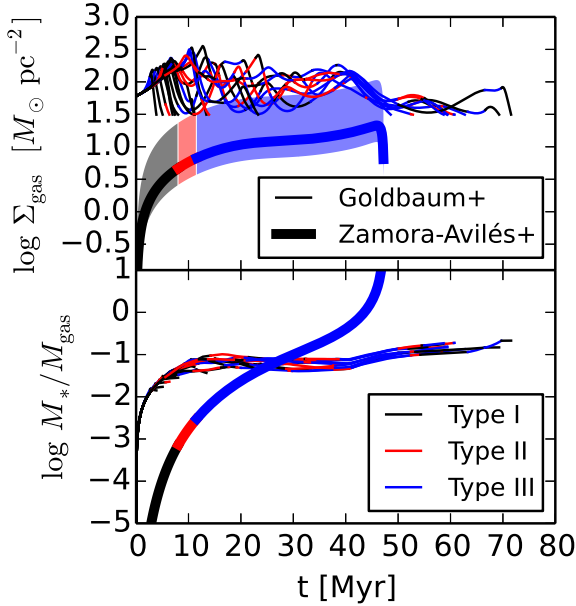}
\caption{\small
\label{fig:gmc_models}
Predictions for the large-scale evolution of GMCs using the models of \citet[thin lines, each line corresponding to a different realization of a stochastic model]{Goldbaum:2011} and \citet[thick line]{Zamora:2012}. The top panel shows the gas surface density. The minimum in the \citeauthor{Goldbaum:2011}~models is the threshold at which CO dissociates. For the planar \citeauthor{Zamora:2012}~model, the thick line is the median and the shaded region is the 10th - 90th percentile range for random orientation. The bottom panel shows the ratio of instantaneous stellar to gas mass. Colors indicate the type following the \citet{Kawamura:2009} classification (see Section 2.5), computed based on the H$\alpha$ and $V$-band luminosities of the stellar populations.
}
\end{figure}

One can also generalize the quasi-stationary turbulent support models by embedding them in time-dependent models for the evolution of a cloud as a whole. In this approach one computes the instantaneous SFR from a cloud's current state (using one of the turbulent support models or based on some other calibration from simulations), but the total mass, mean density, and other quantities evolve in time, so that the instantaneous SFR does too. 
In this type of model, a variety of assumptions are necessarily made about the cloud's geometry and about the effect of the stellar feedback. \citet{Krumholz:2006} and \citet{Goldbaum:2011} adopt a spherical geometry and compute the evolution from the virial theorem, assuming that feedback can drive turbulence that inhibits collapse. As illustrated in Figure \ref{fig:gmc_models}, they find that 
most clouds undergo oscillations around equilibrium before being destroyed at final SFEs $\sim$ 5--10\%. The models match a wide range of observations, including the distributions of column density, linewidth-size relation, and cloud lifetime. In constrast, \citet[also shown in Figure \ref{fig:gmc_models}]{Zamora:2012} adopt a planar geometry \citep[which implies longer free-fall times than in the spherical case;][]{Toala:2012} and assume that feedback does not drive turbulence or inhibit contraction. With these models they reproduce the star formation rates seen in low- and high-mass clouds and clumps, and the stellar age distributions in nearby clusters. 
As shown in the Figure, the overall evolution is quite different in the two models, with the \citeauthor{Goldbaum:2011}~clouds undergoing multiple oscillations at roughly fixed $\Sigma$ and $M_*/M_{\rm gas}$, while the \citeauthor{Zamora:2012}~model predicts a much more monotonic evolution.
Differentiating between these two pictures will require a better understanding of the extent to which feedback is able to inhibit collapse.

\bigskip
\noindent
\textbf{5.2 Connection Between Local and Global Scales}
\bigskip

Extraglactic star formation observations at large scales average over regions several times the disk scale height in width, and over many GMCs. As discussed in Section 2.1, there is an approximately linear correlation between the surface densities of SFR and molecular gas in regions where $\Sigma_{\rm gas} \lesssim 100 M_\odot \ {\rm pc}^{-2}$, likely because observations are simply counting the number of GMCs in a beam. At higher $\Sigma_{\rm gas}$, the volume filling factor of molecular material approaches unity, and the index $N$ of the correlation $\Sigma_{\rm SFR} \propto \Sigma_{\rm gas}^N$ increases. This can be due to increasing density of molecular gas leading to shorter gravitational collapse and star formation timescales, or because higher total gas surface density leads to stronger gravitational instability and thus faster star formation. 
At the low values of $\Sigma_{\rm gas}$ found in the outer disks of spirals (and in dwarfs), the index $N$ is also greater than unity.
This does not necessarily imply that there is a cut-off of $\Sigma_{\rm SFR}$ at low
gas surface densities,  although simple models of gravitational instability in isothermal disks can indeed reproduce this 
    result \citep{Li:2005},
but instead may indicate that additional parameters beyond just $\Sigma_{\rm gas}$ control $\Sigma_{\rm SFR}$.  In outer disks, the ISM is mostly diffuse atomic gas and the radial scale length of $\Sigma_{\rm gas}$ is quite large (comparable to the size of the optical disk; \citet{Bigiel:2012}). The slow fall-off of $\Sigma_{\rm gas}$ with radial distance implies that the sensitivity of $\Sigma_{\rm SFR}$ to other parameters will become more evident in these regions. For example, a higher surface density in the old stellar disk appears to raise $\Sigma_{\rm SFR}$ \citep{Blitz:2004,Blitz:2006,Leroy:2008}, likely because stellar gravity confines the gas disk, raising the density and lowering the dynamical time. Conversely, $\Sigma_{\rm SFR}$ is lower in lower-metallicity galaxies \citep{Bolatto:2011}, likely because  lower dust shielding against UV radiation inhibits the formation of a cold, star-forming phase \citep{Krumholz:2009b}. 

Feedback must certainly be part of this story. Recent large scale simulations of disk galaxies have consistently pointed to the need for feedback to prevent runaway collapse and limit star formation rates to observed levels
\citep[e.g.,][]{Kim:2011,Tasker:2011,Hopkins:2011,Hopkins:2012b,Dobbs:2011,Shetty:2012,Agertz:2013}.  With feedback parameterizations that yield realistic SFRs, other ISM properties (including turbulence levels and gas fractions in different H~\textsc{i} phases) are also realistic \citep[see above and also][]{Joung:2009,Hill:2012}. However, it still also an open question whether feedback is the entire story for the large scale SFR. 
In some simulations \citep[e.g.,][]{Ostriker:2011,Dobbs:2011,Hopkins:2011,Hopkins:2012b,Shetty:2012,Agertz:2013}, the SFR on $\gtrsim 100$ pc scales is mainly set by the time required for gas to become gravitationally-unstable on large scales and by the parameters that control stellar feedback, and is insensitive to the parameterization of star formation on $\lesssim$ pc scales.
In other models the SFR is sensitive to the parameters describing both feedback and small-scale star formation (e.g., $\epsilon_{\rm ff}$ and H$_2$ chemistry; \citealt{Gnedin:2010, Gnedin:2011, Kuhlen:2012, Kuhlen:2013}).

Part of this disagreement is doubtless due to the fact that current simulations do not have sufficient resolution to include the details of feedback, and in many cases they do not even include the required physical mechanisms (for example radiative transfer and ionization chemistry). Instead, they rely on subgrid models for momentum and energy injection by supernovae, radiation, and winds, and the results depend on the details of how these mechanisms are implemented. Resolving the question of whether feedback alone is sufficient to explain the large-scale star formation rate of galaxies will require both refinement of the subgrid feedback models using high resolution simulations, and comparison to observations in a range of environments.
In at least some cases, the small-scale simulations have raised significant doubts about popular subgrid models \citep[e.g.,][]{krumholz:2012b, krumholz:2013}.

A number of authors have also developed analytic models for large-scale star formation rates in galactic disks. \citet{Krumholz:2009b} propose a model in which the fraction of the ISM in a star-forming molecular phase is determined by the balance between photodissociation and H$_2$ formation, and the star formation rate within GMCs is determined by the turbulence-regulated star formation model of \citet{Krumholz:2005}. This model depends on assumed relations between cloud complexes and the properties of the interstellar medium on large scales, including the assumption that the surface density of cloud complexes is proportional to that of the ISM on kpc scales, and that the mass fraction in the 
warm atomic ISM is negligible compared to the mass in cold atomic and molecular phases.  
\citet{Ostriker:2010} and \citet{Ostriker:2011} have developed models in which star formation is self-regulated by feedback. In these models, the equilibrium state is found by simultaneously balancing ISM heating and cooling, turbulent driving and dissipation, and gravitational confinement with pressure support in the {\it diffuse} ISM. The SFR adjusts to a value required to maintain this equilibrium state. Numerical simulations by \citet{Kim:2011} and \citet{Shetty:2012} show that ISM models including turbulent and radiative heating feedback from star formation indeed reach the expected self-regulated equilibrium states. However, as with other large-scale models, these simulations rely on subgrid feedback recipes whose accuracy have yet to be determined. In all of these models, in regions where most of the neutral ISM is in gravitationally bound GMCs, $\Sigma_{SFR}$ depends on the internal state of the clouds through the ensemble average of $\epsilon_{\rm ff}/\tau_{\rm ff}$.  If GMC internal states are relatively independent of their environments, this would yield 
values of $\langle \epsilon_{\rm ff}/\tau_{\rm ff}\rangle$ that do not strongly vary within a galaxy or from one galaxy to another, naturally explaining why $\tau_{\rm dep}({\rm H}_2)$ appears to be relatively uniform, $\sim 2$ Gyr wherever $\Sigma_{\rm gas} \lesssim 100 M_\odot \ {\rm pc}^{-2}$.

Many of the recent advances in understanding large-scale star
formation have been based on disk galaxy systems similar to our own
Milky Way.  Looking to the future, we can hope that the methods being
developed to connect individual star-forming GMCs with the larger scale ISM
in local ``laboratories'' will inform and enable efforts in
high-redshift systems, where conditions are more extreme and
observational constraints are more challenging.

\bigskip
\centerline{\textbf{ 6. Looking forward}}
\bigskip
\noindent
\textbf{6.1 Observations}
\bigskip
\\
\\
Systematic surveys and detailed case studies will be enriched by the expansion in millimeter-submillimeter 
capabilities over the next decade.
These will contribute in two main modes:
through cloud-scale observations in other
galaxies, and in expanding the study of clouds in our own Milky Way.
The increased sensitivities of ALMA and NOEMA
will sample smaller scales at larger distances, resolving GMC complexes and investigating the 
physical state and formation mechanisms of GMCs in a variety of extragalactic environments. 
The smaller interferometers like CARMA and SMA will likely 
focus on systematic mapping of large areas in the Milky Way or even external galaxies. At cm wavelengths, the recently upgraded JVLA brings new powerful capabilities in continuum detection at 7~mm to study cool dust in disks, as well as the study of free-free continuum, molecular emission, and radio recombination lines in our own galaxy and other galaxies. Single-dish
mm-wave facilities equipped with array receivers and continuum cameras
such as the IRAM~30m, NRO~45m, LMT~50m, and the future CCAT facility will enable
fast mapping of large areas in the Milky Way, providing the much
needed large scale context to high resolution studies.
These facilities, along with the APEX and NANTEN2 telescopes, will 
pursue multitransition / multiscale surveys sampling star forming cores and the parent cloud material 
simultaneously and providing valuable diagnostics of gas  kinematics and physical conditions. 

There is hope that some of the new observational capabilities will break the theoretical logjam described in this review. At present, it is not possible for observations to distinguish between the very different cloud formation mechanisms proposed in Section 3, nor between the mechanisms that might be responsible for controlling cloud density and velocity structure, and cloud disruption (Section 4), nor between various models for how star formation is regulated (Section 5). There is reason to hope that the new data that will become available in the next few years will start to rule some of these models out.

\bigskip
\noindent
\textbf{6.2 Simulations and theory}
\bigskip
\\
The developments in numerical simulations of GMCs since PPV will 
continue over the next few years to PPVII. There is a current convergence of simulations towards the scales of GMCs, and GMC scale physics. Galaxy simulations are moving towards ever smaller scales, whilst individual cloud simulations seek to include more realistic initial conditions. At present, galactic models are limited by the resolution required to adequately capture stellar feedback, internal cloud structure, cloud motions and shocks. They also do not currently realise the temperatures or densities of GMCs observed, nor typically include magnetic fields. Smaller scale simulations, on the other hand, miss larger scale dynamics such as spiral shocks, shear and cloud--cloud collisions,
which will be present either during the formation or throughout the evolution of a molecular cloud. Future simulations, which capture the main physics on GMC scales, will provide a clearer picture of the evolution and lifetimes of GMCs. Furthermore they will be more suitable for determining the role of different processes in driving turbulence and regulating star formation in galaxies, in conjunction with analytic models and observations. 

Continuing improvements in codes, including the use of moving mesh codes in addition to AMR and SPH methods, will also promote progress on GMCs descriptions, and allow more consistency checks between different numerical methods. More widespread, and further development of chemodynamical modelling will enable the study of different tracers in cloud and galaxy simulations.  In conjunction with these techniques, synthetic observations, such as HI and CO maps, 
will become increasingly important for comparing the results of numerical models and simulations, and testing whether the simulations are indeed viable representations of galaxies and clouds.
\\
\\
\\
\textbf{ Acknowledgments.} 
Support for this work was provided by: the European Research Council through the FP7 ERC starting grant project LOCALSTAR (CLD); the NSF through grants AST09-55300 (MRK), AST11-09395 (M-MML), AST09-08185 (ECO), AST09-55836 (ADB), and AST10-09049 (MH); NASA through ATP grant NNX13AB84G (MRK), Chandra award number GO2-13162A (MRK) issued by the Chandra X-ray Observatory Center, which is operated by the Smithsonian Astrophysical Observatory for and on behalf of the National Aeronautics Space Administration under contract NAS8-03060, and Hubble Award Number 13256 (MRK) issued by the Space Telescope Science Institute, which is operated by the Association of Universities for Research in Astronomy, Inc., under NASA contract NAS 5-26555; a Research Corporation for Science Advancement Cottrell Scholar Award (ADB); an Alfred P. Sloan Fellowship (MRK); the hospitality of the Aspen Center for Physics, which is supported by the National Science Foundation Grant PHY-1066293 (MRK); and CONACYT grant 102488 (EVS).
\bigskip


\bibliographystyle{ppvi_lim1}
\bibliography{cit}

\end{document}